\documentclass[fleqn,usenatbib]{mnras}
\usepackage{graphicx}
\usepackage{amssymb}
\usepackage{amsmath}
\usepackage{mathptmx}
\usepackage{array,booktabs}
\usepackage[normalem]{ulem}

\newcommand{\approptoinn}[2]{\mathrel{\vcenter{
  \offinterlineskip\halign{\hfil$##$\cr
    #1\propto\cr\noalign{\kern2pt}#1\sim\cr\noalign{\kern-2pt}}}}}

\usepackage{multirow}

\newcolumntype{P}[1]{>{\centering\arraybackslash}p{#1}}
\newcolumntype{M}[1]{>{\centering\arraybackslash}m{#1}}

\graphicspath{{./}{figures/}}

\title[Measuring geometry from off-axis afterglow images]{Analytic Model for Off-Axis GRB Afterglow Images - Geometry Measurement and Implications for Measuring $H_{0}$}
\author[Govreen-Segal \& Nakar]{
    Taya Govreen-Segal\thanks{taya@govreensegal.com}, Ehud Nakar
	\\
	{School of Physics and Astronomy, Tel Aviv University, Tel Aviv 6997801, Israel}
}
\pubyear{2023}

\begin{document}
	\label{firstpage}
	\pagerange{\pageref{firstpage}--\pageref{lastpage}}
	\maketitle

\begin{abstract}
We present an analytic model for measuring the jet core angle ($\theta_c$) and viewing angle ($\theta_{obs}$) of off-axis gamma-ray bursts independently of the jet angular structure outside of the core. We model the images of off-axis jets and using this model we show that $\theta_{obs}$ and $\theta_c$ can be measured using any two of the three following observables: the afterglow light curve, the flux-centroid motion, and the image width. The model is calibrated using 2D relativistic hydrodynamic simulations with a broad range of jet angular structures. We study the systematic errors due to the uncertainty in the jet structure and find that when using the light curve and centroid motion to determine $\theta_{obs}$ and $\theta_c$, our formulae can be accurate to a level of 5-10\% and 30\%, respectively. In light of the Hubble tension, the systematic error in $\cos\theta_{obs}$ in GRBs originating in a binary compact object merger is of special interest. We find that the systematic uncertainty on the measurement of $\cos\theta_{obs}$ due to the unknown jet structure is smaller than $1.5\%$ for well-observed events. A similar error is expected if the microphysical parameters evolve at a level that is not easily detected by the light curve.
Our result implies that this type of systematic uncertainty will not prevent measurement of $H_0$ to a level of $2\%$ with a sample of well-observed GW events with resolved afterglow image motion. Applying our model to the light curve and centroid motion observations of GW170817 we find  $\theta_{obs}=19.2\pm 2~\deg$ (1$\sigma$) and  $\theta_c=1.5-4~\deg$.
\end{abstract}
\section{Introduction} \label{sec:intro}
Merging neutron stars (and most likely also black hole-neutron star mergers) emit gravitational waves (GW) and electromagnetic (EM) radiation throughout the entire electromagnetic spectrum \citep[e.g.,][]{Eichler1989,Li1998,LIGO_multi2017}. The ultra relativistic jets launched in these events probe various aspects of the merger such as the resulting compact object and the accretion disk that surrounds it as well as the sub-relativistic ejecta that the merger throws along the poles \citep[e.g.,][]{Margalit2017,Sarin2021,Nakar2020}. In addition, observations of these jets can also significantly increase the accuracy with which $H_0$ can be measured using the GW-EM signal \citep{Schutz1986,Chen2018,Hotokezaka2019}. These jets are observed, most likely, as short GRBs when observed on axis, and thanks to the alert provided by gravitational waves, can be seen during the afterglow phase by off-axis observers as well \citep{Rhoads1997,Rhoads1999}. 

Off-axis GRBs provide the opportunity to probe properties of GRB jets that are nearly impossible to measure in on-axis GRBs, such as the jet structure, and core angle  \citep[e.g.,][]{Takahashi_Ioka2020,Ryan2020,Mooley2018,Mooley2022,Ghirlanda2019}. The primary reason for this is that while the emission seen from on-axis jets always originates from the jet core, the afterglow of off-axis jets is dominated by emission coming from different parts of the jet at different times, thus encoding various properties of the jet geometry in the light curve \citep{Takahashi_Ioka2020,Ryan2020,Beniamini2020,Beniamini2022}. Off-axis GRBs triggered by GW have the additional advantage of being closer than typical on-axis GRBs and therefore have a bright and observable afterglow for a longer time than the on-axis GRBs that are typically detected.

The geometry of GRBs, that is, the jet core angle,  $\theta_c$, the observing angle, $\theta_{obs}$, and the jet structure, is of special interest. In this work, we focus on the constraints that observations pose on $\theta_{obs}$ and $\theta_c$. The core angle of a GRB jet is of interest as it is related to the total energy in the jet, the jet propagation, and the launching and collimation mechanisms. There have been many attempts to measure the core angle of both short and long GRBs. In on-axis GRBs, the core angle is identified from the jet-break in the light curve, which is challenging to securely identify, and provides a value for the jet core angle that is degenerate with the jet energy and the external density \citep{Rhoads1999,Sari1999}. As demonstrated by GW170817, the geometry of nearby off-axis GRBs can be constrained much more accurately than that of the typically observed on-axis GRBs (\citealt{Mooley2018,Mooley2022,Ghirlanda2019}; see also \citealt{Nakar2020} and references therein). In fact, the jet opening angle and the viewing angle of GW170817 are better constrained than those of any of the thousands of on-axis GRBs that were observed to date. 

The observing angle of off-axis GRBs is also interesting. One application for the observing angle of GW sources is in the context of measuring $H_0$. Gravitational waves provide a measurement of luminosity distance which alongside the host galaxy redshift can be used to measure the Hubble constant \citep{Schutz1986}. However, the luminosity distance is degenerate with the inclination angle, and this degeneracy is expected to be the primary source of uncertainty in $H_0$ in most GW measurements with high signal-to-noise ratios. Assuming that the jets are aligned with the total angular momentum, a measurement of the observing angle with respect to the jet axis can be used to lift the degeneracy and improve the accuracy significantly \citep{Hotokezaka2019}. For such a measurement to be useful, the systematic errors in the measurement of the observing angle must be small and well understood. 

The most attainable observation of GRB afterglows, the light curve, plays a major role in constraining the jet geometry and the observing angle. In an off-axis afterglow observed at a frequency that is above the self-absorption and typical synchrotron frequencies, the rising part of the light curve probes the angular structure of the jet outside of the core. In their studies of the rising phase, \cite{Ryan2020} and \cite{Takahashi_Ioka2020} show that the light curve does not provide a unique solution for the jet structure. In fact, they show that while the rising phase teaches us a lot about the jet structure, there is still enough freedom so that infinitely many structures can fit an observed light curve. \cite{Nakar2021} have shown that the width of the peak of the light curve provides a measurement of the ratio between the viewing angle and the jet core angle $\theta_{obs}/\theta_c$. They show that a single formula (relating the peak width to $\theta_{obs}/\theta_c$), which is independent of the jet structure, is expected to provide a good estimate for this ratio. They use a semi-analytic model to calibrate this formula and they do not quantify its accuracy for various jet structures. Finally, the declining part of the light curve does not provide useful constraints on the jet geometry, at least as long as the jet is relativistic, since the light curve during this phase joins the one seen by an on-axis observer \citep{Granot2002,Nakar2002}.

\cite{Nakar2021} conclude that the light curve alone can only be used to measure $\theta_{obs}/\theta_c$ and additional information is needed in order to measure each of these angles separately. Observing the motion of the afterglow image on the plane of the sky provides such information. \cite{Mooley2018} demonstrated with the GRB afterglow of GW170817 that two VLBI radio images of the afterglow measured at around the time of the peak of the light curve can be used to measure $\theta_{obs}-\theta_c$ and break the degeneracy. They used a limited number of hydrodynamic simulations to obtain a simultaneous fit to the light curve and the centroid motion, thereby constraining $\theta_{obs}$ and $\theta_c$ separately. This method was supported by using an approximate analytic model of the image centroid motion alongside the light curve to constrain the angles. Later, \cite{Ghirlanda2019} obtained a measurement of the centroid location at a third epoch near the peak. They derived constraints on $\theta_{obs}$ and $\theta_c$ based on a simultaneous fit for the light curve and centroid motion to semi-analytical models of non-spreading jets. Finally, \cite{Mooley2022} presented an additional astrometric measurement of the optical kilonova using the Hubble Space Telescope, which provides the location of the jet origin. Using this observation along with the radio image observations from \cite{Mooley2018} and \cite{Ghirlanda2019}, they measure $\theta_{obs}$ and $\theta_c$ by fitting the light curve and centroid-motion simultaneously to a set of approximated hydrodynamical simulations. 
All these studies presented a significant improvement over previous works using only the afterglow light curve. However, the approximations they used limit the accuracy of the results, and more importantly, they do not account for the uncertainty that arises from the unknown jet angular structure. 

The first goal of this work is to provide numerically calibrated analytic formulae that can be used to measure $\theta_{obs}$ and $\theta_c$ independently of the jet structure based on any combination of two of the following observations: the light curve peak width, a measurement of the afterglow image centroid motion, and the afterglow image width (length in the direction perpendicular to the centroid motion). These formulae can be useful in design and analysis of future observations. The second goal is to place an upper limit on the systematic errors that the unknown jet structure introduces to the measurements of $\theta_{obs}$ and $\theta_c$. In this regard we focus on the use of the light curve and centroid motion, since it is unlikely the width of the image will be measured to a better accuracy than the centroid motion. We do this by simulating the images and light curves of jets with a large range of different structures, and then comparing the actual values of $\theta_{obs}$ and $\theta_c$ in the simulations, to the values of $\theta_{obs}$ and $\theta_c$ found by applying the structure-independent analytic formulae to the simulated data. The last goal of this paper is to characterize the shape of the image (its width, depth and general shape). This can increase the sensitivity of VLBI measurements, which currently usually fit the data to an image with some generic shape such as a 2D Gaussian.

Our first step to obtain these goals is to develop an analytic model of the temporal evolution of afterglow images of off-axis jets, and verify and calibrate it with a set of 2D numerical simulations with varied initial jet structures. To date, very little work has been done on the image of off-axis GRBs and no analytic model for GRB afterglow images from off-axis jets exists. \cite{Sari1998} and \cite{Granot1999} modeled analytically the image of on-axis GRBs while the observed region is still quasi-spherical. \cite{Gill2018} used a semi-analytical model to study the afterglow images and polarization of several outflows with a light curve that fit the observations of GW170817, and concluded that measurement of the centroid motion or of linear polarization can be used to distinguish a jet from a more spherical outflow. \cite{Granot2018} and \cite{Zrake2018} presented images of off-axis afterglows based on relativistic hydrodynamic simulations from several specific setups, without deriving analytic formulae that relate the jet properties and the observed image. Finally, \cite{Fernandez2022} used various semi-analytic models to highlight the importance of jet spreading on the image centroid motion of off-axis jets  in these models.

Our next step is calibrating the model from \cite{Nakar2021} for finding $\theta_{obs}/\theta_c$ from the shape of the peak of the light curve, using the set of 2D jet simulations. We then combine the two calibrated models to provide analytic (numerically calibrated) expressions for $\theta_{obs}$ and $\theta_c$ that depend on the image centroid motion and on the width of the peak of the light curve. We also show that the image width can be used either alongside the light curve observations or along measurements of the centroid motion to find $\theta_{obs}$ and $\theta_c$.
Finally, we use our model and simulations to study the systematic errors one may expect in $\theta_{obs}$, $\theta_c$ and $\cos(\theta_{obs})$, focusing on the systematic errors expected due to uncertainty in the jet structure. As an additional source of systematic errors, we consider the effect of non-constant microphysical parameters ($\epsilon_e,\epsilon_B$), on the inferred values of $\theta_{obs}$ and $\theta_c$, for variations in the microphysical parameters which are not easily identifiable by the light curve.

We proceed as follows. In \S\ref{sec:Analytic}, we develop an analytic model of the off-axis jet afterglow image evolution. We follow, in \S\ref{sec:Numerical_Simulations} by presenting our numerical simulations. In \S\ref{sec:Numarical_Results} we present the numerical results, calibrate the analytic model, and discuss the expected systematic errors. The main results of this work are summarized in \S\ref{sec:core and obs}. In this section we summarize the method for finding $\theta_{obs}$ and $\theta_{c}$, discuss what to consider when planning observations for future events, and apply our model to GW170817. We follow in \S\ref{sec:H0} by discussing the potential for using afterglow image observations to constrain $H_0$ with GW-EM events with observable jet emission. In \S\ref{sec:micro_phys} we consider the effect of non-constant microphysical parameters on the systematic errors of our model. We draw our conclusions in \S\ref{sec:Conclusions}. 

\section{Analytic model}\label{sec:Analytic}
We start by presenting an analytic model for the afterglow of off-axis jets, focusing only on properties of the afterglow that are useful for measuring the viewing angle and the jet opening angle. Our model is derived using various approximations and coefficients of order unity, which are later tested and calibrated using numerical simulations. Below, we describe first the general picture we consider, followed by  a summary of the relevant properties of the light curve, which were derived in previous studies. Then we derive a model for the afterglow image.

\subsection{Model Description \label{sec:model}}
Consider an axisymmetric jet consisting of a core with opening angle  $\theta_{c}$, in which the isotropic equivalent energy, $E_{iso}$, is constant, surrounded by an energy profile that decreases monotonically with the angle from the symmetry axis, $\theta$. The jet is propagating in an external medium with a density profile $\rho \propto r^{-k}, k<4$. We assume the complete jet structure is past the deceleration radius, and neglect the effect of spreading. Thus, we approximate the evolution of the shock at each angle as a part of a spherical explosion in the self-similar phase with  energy $E_{iso}(\theta)\equiv 4\pi \frac{dE(\theta)}{d\Omega}$, where $dE(\theta)$ is the actual energy carried by the jet within a solid angle $d\Omega$. The shock Lorentz factor, $\Gamma$, is therefore completely determined by $E_{iso},n,t$ and evolves as $\Gamma(\theta)\propto \left(E_{iso}(\theta)\right)^{1/2}t^{-\omega}, \omega=\frac{3-k}{2}$ \citep{Blandford1976}, with a proportionality constant that depends only on the density.  For a uniform external density, which will be the focus of this work, $\omega=\frac{3}{2}$. The non-spreading approximation is valid in the jet core at least until the shock front along the core becomes causally connected, namely, the Lorentz factor in the core approaches  $\sim \frac{1}{\theta_c}$. When comparing our analytical model based on this assumption to hydrodynamic simulations (which naturally includes a complete treatment of spreading), in \S\ref{sec:Numarical_Results}, we find  that for some purposes it is a useful approximation also at later times, while for others it is less useful. In places where jet spreading must be taken into account, a useful approximation is that of maximal spreading \citep{Sari1999}, where $\Gamma$ decreases exponentially in the shock radius, and for many purposes the shock radius can be approximated as constant. 

For a given jet structure, the Lorentz factor of the shock depends on the lab time $t$ and the angle $\theta$, namely $\Gamma=\Gamma(t,\theta)$. For brevity, everywhere in the paper, $\Gamma$ is used to denote the local shock Lorentz factor of the point discussed (without reminding the reader that $\Gamma=\Gamma(t,\theta)$).  Specifically, when $\Gamma$ is mentioned alongside an angle, they both refer to the same point in time and space. 

We use the approximation that the radiation is emitted from a thin shell behind the shock according to the standard afterglow model (e.g., \citealt{Sari_piran_Narayan1998}). We assume constant microphysical parameters, and an electron distribution $\frac{d N(\gamma_{e})}{d\gamma_e}\propto\gamma_e^{-p}$, where $\gamma_e$ is the Lorentz factor of the accelerated electrons. As may be expected for radio observations of off-axis jets, we consider observed frequencies that are above the synchrotron self-absorption and typical synchrotron frequencies, $\nu_a$ and $\nu_m$, and below the cooling frequency, $\nu_c$. 

We only consider jet structures and viewing angles for which the light curve has an identifiable peak caused by the jet geometry \citep{Beniamini2020,Beniamini2022}. For most cases, a decay of the energy outside the core - $\frac{d\log E_{iso}}{d\log\theta}(\theta\ge\theta_c)\lesssim -2$, and an observing angle $\theta_{obs}\ge2\theta_c$ are sufficient conditions. In our model, we assume the jet is ultra-relativistic during the peak of the light curve, corresponding to $\theta_{obs}\lesssim 0.5$ rad. In the numerical section, we compare the analytic model also to simulations with viewing angles up to $1$ rad, and find very good agreement until $\theta_{obs}=0.75$ rad, and reasonable agreement until $\theta_{obs}=1$ rad. 

\subsection{Coordinate Systems}
The jet and the afterglow emission may be described in several coordinate systems. Above, we used a spherical coordinate system $\left(r,\theta,\phi\right)$, oriented with the pole, $\theta=0$, along the jet axis, and with the origin at the jet origin. When considering an off-axis jet, it is convenient to work in  a coordinate system aligned with the observer rather than the jet; $(r,\theta,\phi)\to(r,\xi,\psi)$ where in the new coordinate system, the observer is at the pole, $\xi$ is the polar angle and $\psi$ is the azimuthal angle, chosen so that the jet axis is at $\psi=0$. 
The coordinate transformation is given by: 
 \begin{equation} \label{eq:cosxi}
 \cos\xi = \cos\theta_{obs}\cos\theta+\sin \theta_{obs}\sin\theta \cos\phi
 \end{equation}
 \begin{equation}\label{eq:cospsi}
 \cos \psi = \frac{\cos\theta-\cos\theta_{obs}\cos\xi}{\sin\xi\sin \theta_{obs}}
 \end{equation}
 Note that for $\phi=0$ Eq. \eqref{eq:cosxi} reduces to: $\xi=\theta_{obs}-\theta$. (This can be seen also in Fig. \ref{fig:ring_jet_intersection}). Both these coordinate systems are accompanied by a lab time $t$ where $t=0$ at the jet launching time.
 
 To write the equations that govern the shape of the image, we must define a third coordinate system - the 2D sky coordinate system onto which the image is projected - $(x,y)$. The sky coordinates are chosen such that the origin is on the line-of-sight, i.e., $\xi=0$ and the $y$ axis is the projection of the jet axis, $\theta=0$, so the jet image centroid starts at the origin and advances in the positive $y$ direction (at least as long as the jet is ultra-relativistic). 
A photon emitted at some point $(r, \xi,\psi)$ will be observed on the sky at:
 \begin{equation}
     x = r\sin\xi \sin \psi\simeq r\xi \sin \psi
 \end{equation}  \begin{equation}\label{eq:y_obs}
     y = r\sin \xi \cos \psi \simeq r\xi \cos \psi~~ ,
 \end{equation}
where we approximated $\xi\ll1$, an approximation that is valid at least while the emitting region is ultra-relativistic, and that we will use for the rest of the paper.
 The radius of such a point in the sky coordinates is:
 \begin{equation}\label{eq:R_sky}
 R = \sqrt{x^2+y^2}=r\sin \xi\simeq r\xi  
 \end{equation}
 And the emitted photon will be observed at an observer time $T$:
 \begin{equation}\label{eq:T}
T=t-r\cos \xi/c.
\end{equation}
The notations used in this work are summarized in the glossary in table \ref{tab:glossary}.

\begin{table*}
    \centering
\begin{tabular}{|c|c|}
\hline 
Symbol & Definition\tabularnewline
\hline 
\hline 
$\left(r,\theta,\phi\right)$ & Lab-frame spherical coordinate system,  origin at the jet source, oriented with the jet axis
at the pole $\theta=0$\tabularnewline
\hline 
$\left(r,\xi,\psi\right)$ & Lab-frame spherical coordinate system, origin at the jet source, oriented with the observer
at the pole $\xi=0$\tabularnewline
\hline 
$t$ & Lab time\tabularnewline
\hline 
$\left(x,y\right)$ & Plane of the sky coordinate system; the origin is at the line-of-sight, jet axis is projected on the $y$ axis\tabularnewline
\hline 
$R$ & $=\sqrt{x^{2}+y^{2}}$; radius in the sky coordinate system\tabularnewline
\hline 
$T$ & Observer time\tabularnewline
\hline 
$\theta_{c}$ & Jet core angle; in the numerical section, this is the jet core angle
at $T_p$ as defined by Eq. \eqref{eq:core_def}\tabularnewline
\hline 
$b$ & Power-law index of power-law jets; $E_{iso}\left(\theta>\theta_{c}\right)\propto\theta^{-b}$\tabularnewline
\hline 
$\theta_{obs}$ & Observer angle (measured from the jet axis)\tabularnewline
\hline 
$E_{iso}$ & $=4\pi\frac{dE}{d\Omega}$; local isotropic equivalent energy (depends on $\theta$) \tabularnewline
\hline 
$\Gamma$ & Local shock Lorentz factor at the point of interest 
 \tabularnewline
 & (i.e; $\Gamma\xi=1$
means $\Gamma$ and $\xi$ at the same point on the jet, selected
so that their product is 1)\tabularnewline
\hline 
$k$ & External medium density profile index, $\rho\propto r^{-k},k<4$.
In this work we focus on $k=0$. \tabularnewline
\hline 
$\omega$ & $\Gamma\propto t^{-\omega}$; for a spherical explosion, $\omega=\frac{3-k}{2}$\tabularnewline
\hline 
$p$ & Electron distribution power-law index \tabularnewline
\hline 
$T_{p}$ & The light curve peak time \tabularnewline
\hline 
$T_{end}$ & The time at which $\frac{d\log F_{\nu}}{d\log T}=-p$ for the first
time \tabularnewline
\hline 
$h$ & Between $T_p$ and $T_{end}$ $\Gamma \propto T^{-h}, \frac{3}{8}\le h\le\frac{1}{2}$\tabularnewline
\hline 
$R_{arc}$ & The radius of the arc of the image on the plane of the sky\tabularnewline
\hline 
$y_{cen}$ & The flux weighted centroid of the image (always on the y-axis)\tabularnewline
\hline 
$y_{max},y^{max}$ & Largest $y$ of the image\tabularnewline
\hline 
$\xi^{max}$ & $\xi$ for which $(\xi,\psi)=(\xi^{max},0)$ contributes to the emission at $y_{max}$\tabularnewline
\hline 
$\Delta y$ & Image depth (the length of the image when projected onto the $y$
axis)\tabularnewline
\hline 
$\Delta x$ & Image width (the length of the image from smallest to largest
$x$)\tabularnewline
\hline 
\end{tabular}
    \caption{A glossary of the notation used in this paper.}
    \label{tab:glossary}
\end{table*}
\subsection{Light Curve}

The observed light curve is determined by the energy profile of the jet, the external medium density, the microphysics and the geometry of the system. The light curve is expected to show four phases\footnote{In some cases there may be an additional phase preceding the phases we consider in our paper \citep{Beniamini2020,Beniamini2022}. If the viewing angle and jet structure are such that before the jet decelerates, the matter directly on the observers line of sight dominates the emission, than the light curve will rise as the shock collects matter, and have a first peak when the shock starts decelerating, followed by a decrease in the light curve similar to that seen by an on-axis observer, as the matter on the line of sight decelerates. During this time the jet structure is unimportant. The rising phase that we consider in this paper starts then when the jet material that is closer to the core and away from the line-of-sight starts dominating the observed emission. Thus, if the light curve shows two peaks, then here we consider the emission starting at the rising phase of the second peak.} - a rise, a peak, a steep decay and a shallow decay. The shape of the light curve in each of the phases depends
on different parameters. The shape of the rising phase depends on the structure of the jet outside of its core \citep{Takahashi_Ioka2020, Ryan2020}. The shape of the peak of the light curve depends on the ratio $\frac{\theta_{obs}}{\theta_c}$ and on $p$ \citep{Nakar2021}, and is only weakly dependent on the jet structure. The decaying phases are power-laws with indices that depend on $p$. The overall normalization depends on all of the parameters and the transition times between the phases depend on the system geometry and on the ratio between the energy and the external density \citep{Granot2002,Nakar2002}.

As long as the jetted blast wave is relativistic, the difference in light-travel time to the observer between different parts of the jet means that radiation reaching the observer simultaneously can be traced back to a range of different lab times and corresponding Lorentz factors. However, not all these times contribute equally, in fact, since the emitted radiation is highly beamed, the points dominating the observed flux are those for which the angle to the observers line of sight is $\sim\frac{1}{\Gamma}$ (This is explained in more detail in \S\ref{sec:sphere}). As the jet decelerates, the region dominating the radiation scans through the jet structure, enabling us to use the light curve as a probe of the jet geometry. (see animation \url{youtu.be/WSp-P3kyaoA}).

During the rising phase, the deceleration of the shock wave dictates that following the matter at an angle of $\xi \sim\frac{1}{\Gamma}$ from the line of sight, the observer sees farther along the jet towards the axis. The exact angle dominating the emission and the rate at which that angle travels through the structure depend on the jet structure, and determine the shape of the light curve during the rising phase.  

Once the jet has decelerated enough that the point dominating the flux is at an angle of $\xi \simeq \frac{1}{\Gamma} \simeq \theta_{obs}-\theta_c$ from the line of sight, the light curve peaks and the peak phase starts. At this phase the emission is dominated by all the points in the core with an angle of approximately $\frac{1}{\Gamma}$ from the line of sight. The contribution from matter outside of the core is negligible during this phase. This stage ends at time $T_{end}$, roughly when the observer sees the far side of the jet core, namely, when the emission is dominated by the shocked fluid elements that satisfy $\xi \simeq \frac{1}{\Gamma}\simeq \theta_{obs}+\theta_c$. After this time, the $1/\Gamma$ emission cone from every point within the jet core includes the observer, so the light curve approaches the one seen by an observer with $\theta_{obs}<\theta_{c}$, a so-called on-axis observer. Thus, the light curve slope approaches the asymptotic post jet-break decline. The exact decline rate depends on the details of the jet lateral spreading. For an exponentially spreading jet, and as we find in our simulations, asymptotically $F_{\nu}(T_{end}\le T\le T_{sub-relativistic})\sim T^{-p}$ \citep{Sari1999}. Geometrical effects cause the light curve to overshoot, initially declining more rapidly than $T^{-p}$ before it approaches its asymptotic value (See \citealt{Granot2007} for a discussion). 
Once the shock becomes sub-relativistic, at $T_{sub-relativistic}$ , the light curve decline becomes shallower and may even show a small bump as the counter-jet emission becomes observable. This phase is of no interest to us in this work.

\subsection{Finding $\theta_{obs}/\theta_c$ From the light curve} \label{sec:angles_lightcurve}
From the discussion above we see that the only phase of the light curve from which $\theta_{obs}$ and/or $\theta_c$ can be constrained is 
the peak. The shape of the peak and specifically the ratio $T_{end}/T_{p}$ depends on the ratio $\theta_{obs}/\theta_c$ 
and $p$, with a weak dependence on the jet structure outside of the core. \cite{Nakar2021} derived an approximate analytic expression that relates $T_{end}/T_{p}$ and $\theta_{obs}/\theta_c$. Here we repeat the analytic derivation, since it is brief and useful, and in \S\ref{sec:lightcurve} we test and calibrate it numerically. In order to have an accurate and measurable definition of $T_{end}$, we follow the definition of \cite{Nakar2021}, defining $T_{end}$ as the first time the light curve declines as $T^{-p}$:
\begin{equation}\label{eq:Tend}
T_{end}=T\left(\frac{d\log{F_{\nu}}}{d\log T}=-p\right)
\end{equation}
Between $T_{p}$ and $T_{end}$ we see farther and farther into the core. To find an analytic expression for $T_{end}/T_p$ we approximate the evolution of the Lorentz factor of the observed matter as a power-law with the observer time $\Gamma
\propto T^{-h}$. Under the approximation of a maximal spreading $h=\frac{1}{2}$ (which corresponds to $\omega \longrightarrow \infty$), while the approximation of no spreading implies $h=\frac{3}{8}$ (corresponding to $\omega=3/2$). Recalling that the observed region satisfies $\Gamma(T_p) \simeq (\theta_{obs}-\theta_c)^{-1}$ and  $\Gamma(T_{end}) \simeq (\theta_{obs}+\theta_c)^{-1}$, then under this assumption one obtains \citep{Nakar2021}:
\begin{equation}\label{eq:Tend_T_p_ratio}
    \frac{T_{end}}{T_{p}} \approx  \left(\frac{\Gamma(T_{end})}{\Gamma(T_p)}\right)^{-1/h}\approx C_{end} \left(\frac{\theta_{obs}/\theta_{c}+1}{\theta_{obs}/\theta_{c}-1}\right)^{1/h}
\end{equation}
Where $C_{end}$ and $h$ are calibration constants, which may depend on $p$. The best fit values of $C_{end}$ and $h$ are derived in \S\ref{sec:lightcurve} and given in table \ref{tab:calibration}. It can be seen, as expected, that the best fit value of $C_{end}$ is of order unity and the best fit value of $h$ is in the range $3/8\le h<1/2$.

\subsection{Afterglow Image of an Expanding Sphere}\label{sec:sphere}

Before deriving the image of off-axis jets we summarize known features of the image of an expanding sphere \citep{Sari1998, Granot1999}.
Consider a setup like the one described in \S\ref{sec:model}, except the jet is replaced by a spherical shock propagating in a medium with a uniform density. We are interested in the properties of the image seen by a stationary distant observer. Such an observer will simultaneously receive photons emitted from different shock radii. Consider two photons emitted at a given radius. The one at a larger angle from the observers line of sight will have a longer distance to travel to the observer, and contribute to a later observer time. Thus, at a given observer time, photons that were emitted from larger observing angles carry radiation from smaller shock radii and accordingly, higher shock Lorenz factors. The locus of points from which photons reaches the observer at the same time is the equal arrival time surface (EATS), and the observed image is the projection of the EATS on to the sky. The widest part of the EATS is the region where the angle to the observers line of sight satisfies $\xi \simeq \frac{1}{\Gamma}$. These points are projected to the edge of the image, setting its size (see a detailed mathematical definition of the EATS and its projection on the plane of the sky in \citealt{Sari1998}).

\begin{figure}
    \centering
    \includegraphics[width = \columnwidth]{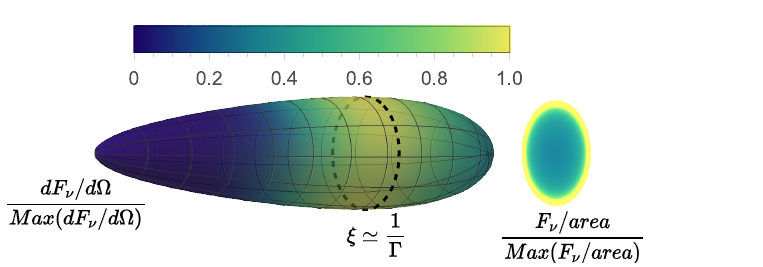}
    \caption{The equal arrival time surface, colored according to the surface brightness, and the image formed by its projection. The dashed line marks the points upon the EATS with the largest cylindrical radius. These points are at an angle of $\xi\simeq \frac{1}{\Gamma}$ where $\Gamma$ is the local Lorentz factor, and $\xi$ is the lab frame angle of the velocity of these points with respect to the line of sight. Points with $\Gamma\xi\simeq 1$ are mapped to the edge of the image. In the sketch, the Lorentz factor along $\xi=0$ is $\Gamma=3$. 
    \label{fig:EATS}}
\end{figure}
Fig. \ref{fig:EATS} depicts the EATS and the image formed by its projection.
The image is fainter in the center and bright along a narrow ring close to its outer edge, where the contributions are from $\Gamma\xi\simeq 1$. As seen in Fig. \ref{fig:EATS}, the apparent bright ring appears is both because the emission along the EATS in the vicinity of $\Gamma\xi=1$ is brighter, and because the shape of the EATS projects a relatively large range of angles from around $\xi \simeq \frac{1}{\Gamma}$ to a small section near the edge of the image on the sky plane, creating relativistic limb-brightening. 

To summarize, the image of an afterglow from a relativistic spherical blast wave  can be described by a bright ring encircling a fainter region. The radiation in the ring comes from points on the sphere with an angle of $\xi \simeq \frac{1}{\Gamma}$. The brightest point in the ring is near the image edge, and the ring is much narrower than the image radius. For example,  \cite{Granot1999} show that for frequencies that satisfy $\nu_a,\nu_m<\nu<\nu_c$ and $p=2.5$, the brightest point in the ring is at $95\%$ of the image radius, and the width of the ring (defined as the distance between the points with half the maximal flux) is $18\%$ of the image radius. The numerical values defining the ring shape are weakly dependent on $p$.

\subsection{Afterglow images of Off-Axis Jets}\label{sec:Image_theory}

We start by considering the image of a top-hat jet, and follow by discussing the effect of a more general angular structure. Since the image shape before the peak of the light curve ($T<T_{p}$) depends on the jet structure we derive only the location of the centroid during this phase. After the peak we derive also width and the depth of the image, however, numerical simulations show that the same formulae can be applied to a range of times before the peak as well. For our derivation we use the relation between the lab time and observer time (Eq. \ref{eq:T}) and the relation between the shock radius and the lab time to express the shock radius in terms of the Lorentz factor at the time of emission, and the observer time $T$ (see \citealt{Sari1998} for a full derivation):
 \begin{equation}\label{eq:R}
r=\frac{c T}{\frac{1}{1-\frac{1}{2\left(2\omega+1\right)\Gamma^{2}}}-\cos\xi}\simeq \frac{c T}{\frac{\xi^{2}}{2}+\frac{1}{2\left(2\omega+1\right)\Gamma^{2}}} 
 \end{equation}

\subsubsection{The Image of a Top-Hat Jet at $T_p<T<T_{end}$}
\begin{figure}
    \centering
\includegraphics[width = \columnwidth]{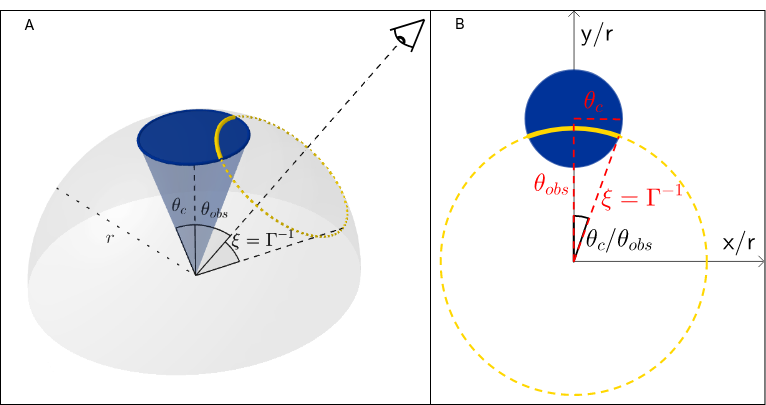}
    \caption{A sketch of the image formed by a top-hat jet - the intersection of the points at an angle $1/\Gamma$ from the observers line of sight (in yellow) and the jet (in blue) forms an arc.}
\label{fig:ring_jet_intersection}
\end{figure}

Consider a top-hat jet, modeled as described in \S\ref{sec:model}, with no energy outside the jet core. 
When the light curve peaks, the observed emission is dominated by the edge of the jet core $(\xi,\psi)\simeq (\theta_{obs}-\theta_c,0)$ where the instantaneous Lorentz factor at the jet edge satisfies $\Gamma(T_p)\simeq \frac{1}{\theta_{obs}-\theta_{c}}$.  From $T_p$ we start seeing into the core. At $T_p<T<T_{end}$ the observed image is a section of the bright ring we would have observed if the blast-wave would have been a complete sphere. Thus, the image in this case is a bright arc, formed by the intersection of the circle formed by all the points with an angle of  $\xi \approx \frac{1}{\Gamma}$ to the observer and the circular region $\theta \leq \theta_{c}$ around the jet axis (see schematic sketch in Fig. \ref{fig:ring_jet_intersection}).
To parameterize this arc, we must find its radius, angle and typical thickness, than translate them into potentially observable quantities -  centroid location $y_{cen}$ (which due to symmetry is along the $y$ axis), width $\Delta x$, and depth $\Delta y$. 

The radius of the arc on the plane of the sky at the time of the peak is given by the radius of the point $(\xi,\psi)=(\frac{1}{\theta_{obs}-\theta_c},0)$. This point has a Lorentz factor $\Gamma\simeq\frac{1}{\theta_{obs}-\theta_{c}}$, and plugging it into equations \eqref{eq:R_sky} and \eqref{eq:R}, we find: 
\begin{equation}
 R_{arc}(T=T_p)\simeq\frac{1+2\omega}{1+\omega}\frac{c T_{p}}{\theta_{obs}-\theta_{c}} ~.
\end{equation}

This can be generalized to any time between $T_p$ and $T_{end}$.
For any location within the core Eq. \ref{eq:R} dictates that the relation between the Lorentz factor and the observer time is given by:
 \begin{equation}\label{eq:T(Gamma)}
T=T_{0} \left(\frac{\Gamma}{\Gamma_{0}}\right)^{-1/\omega} \left(\frac{\xi^{2}+\frac{1}{\left(2\omega+1\right)\Gamma^{2}}}{\xi_{0}^{2}+\frac{1}{\left(2\omega+1\right)\Gamma_{0}^{2}}}\right)
 \end{equation} where $\Gamma_0$ is the shock Lorentz factor of the matter at $\xi_{0}$ contributing to observer time $T_0$.
 From Eq. \eqref{eq:T(Gamma)} and the approximation that $\Gamma$ and $\xi$ at the points dominating the emission obey  $\Gamma\xi=1$, one can show that
$\frac{T}{T_p}=\left(\frac{\xi}{\theta_{obs}-\theta_{c}}\right)^\frac{2\omega+1}{\omega}$ and hence:
\begin{equation}
 R_{arc}\left(T_{p}\le T\le T_{end}\right)=\frac{1+2\omega}{1+\omega}\frac{c T}{\theta_{obs}-\theta_{c}} \left(\frac{T}{T_p}\right)^{-\frac{\omega}{2\omega+1}}~.
 \end{equation}

The half opening angle of the arc, at least for a part of the evolution, can be found  by simple geometry (see Fig. \ref{fig:ring_jet_intersection}) to be $\simeq\frac{\theta_c}{\theta_{obs}}$. 
From the arc radius and angle and the fact that the depth of the image is thin compared to the arc radius, we can use $y_{cen}\approx R_{arc}$ (see quantification below), and we find that the width of the image is
\begin{equation}\label{eq: width}
\Delta x \simeq 2\frac{\theta_c}{\theta_{obs}} y_{cen}~.
\end{equation}
We numerically find this to be a reasonable approximation for most jet structures between $0.5T_p$ and $T_{end}$. A calibration coefficient for this relation  (found to be of order unity) is given in appendix \ref{Appendix:depth}.
The depth of the image, $\Delta y=\max (y)-\min (y)$ is less straightforward to derive, since it depends on the shape formed by the intersection of the ring and the core. In appendix \ref{Appendix:depth} we give analytical arguments for the depth of the image, and find that for $\theta_{obs}/\theta_c>2$,  $\frac{\Delta y}{\Delta x}\le1$, and the ratio approaches unity for for $\theta_c\ll \theta_{obs}$. We also find that the typical value of the width is $\Delta y/R_{arc}\simeq 0.2-0.3$ (see Fig. \ref{fig:depth_over_width}).


Considering the centroid location, since $\Delta y\ll R_{arc}$, we can approximate $y_{cen}\simeq R_{arc}$, and introduce a calibration constant to calibrate the relation:
\begin{equation}\label{eq:centroid_a}
y_{cen}=C_{cen} R_{arc}~~~;~~~T_p\le T\le T_{end}
\end{equation}
Where we expect $C_{cen}\simeq 1$. This calibration constant also accounts for the actual value of $\Gamma\xi$ dominating the emission at the time of the peak (which is not exactly unity, see discussion in appendix \ref{Appendix:Gamma_xi}). 

\subsubsection{The Image Centroid Motion of a Top-Hat Jet at $T<T_p$ }
Before the peak of the light curve, $\Gamma\ge \frac{1}{\theta_{obs}-\theta_c}$ in the entire jet and no part of its emission is beamed towards the observer. The brightest point is the point at the smallest angle to the line of sight - $\xi =\theta_{obs}-\theta_{c}$. This point also sets $y_{max}$; the maximal $y$ of the image. Setting $(\xi,\psi) =(\theta_{obs}-\theta_{c},0)$ in equations \eqref{eq:y_obs} and \eqref{eq:R}, we find that the location of this point on the sky is given by:
\begin{equation}\label{eq:yobs_bp}
y_{max}=\frac{2 c T (\theta_{obs}-\theta_c) }{\frac{1}{\Gamma ^2 (2 \omega +1)}+(\theta_{obs}-\theta_c) ^2} ~~,
\end{equation}
where $\Gamma$ is the Lorentz factor at $\xi = \theta_{obs}-\theta_c$. The relation between the Lorentz factor and the observer time can be found by plugging $\xi=\theta_{obs}-\theta_c$ in Eq. \eqref{eq:T(Gamma)}, and setting the Lorentz factor at the time of the peak as $\Gamma(T_{p})=\frac{1}{\theta_{obs}-\theta_c}$:
\begin{equation}\label{eq:T_bp}
\frac{T}{T_{p}}=\frac{1+\Gamma^{2}\left(\theta_{obs}-\theta_{c}\right)^{2}\left(2\omega+1\right)}{\Gamma^{2+\frac{1}{\omega}}\left(\theta_{obs}-\theta_{c}\right)^{2+\frac{1}{\omega}}\left(2\omega+2\right)}
\end{equation}
Equations \eqref{eq:yobs_bp} and \eqref{eq:T_bp} give a parametric solution for $y_{max}(T\le T_p)$, in terms of $\Gamma$. For $\omega=\frac{3}{2}$, this solution can be approximated with at most $1\%$ error by: 
\begin{equation}
y_{max}\simeq \frac{2 c T}{\left(\theta_{obs}-\theta_c\right)\cdot\left(1+ \frac{1}{4}\left(\frac{T}{T_{p}}\right)^2\right)}~~~;
~~~ T<T_p~.
\end{equation}
Since the beaming depends strongly on the angle, contributions from larger angles should be significantly dimmer, and the whole bright region, as well as the centroid, should be close to $y_{max}$. We can therefore approximate
\begin{equation}\label{eq:centroid_b}
y_{cen}=C_{cen}  \frac{2 c T}{\left(\theta_{obs}-\theta_c\right)\cdot\left(1+ \frac{1}{4}\left(\frac{T}{T_{p}}\right)^2\right)}~~~;
~~~ T<T_p ~,
\end{equation}
where $C_{cen}$ is a calibration coefficient which we find numerically. Note that continuity requires this to be the same constant as the one after the peak, in Eq. \eqref{eq:centroid_a}.

\subsubsection{Afterglow image and centroid motion of a jet with an angular structure} \label{sec:structured_jets}
We expect that the image of a jet with an angular structure will differ from that of a top-hat jet in several ways.
At $T_p<T<T_{end}$, while we see into the core, the contribution from the structure is expected to  extend the arc that we would have seen from a top-hat jet. However, since the emission during this time from $\theta>\theta_c$ is much fainter than the emission of the core, we expect the structure to  extend the arc only slightly. This will have a mild effect on $\Delta y$, a minor effect on $\Delta x$ and an even smaller effect on the centroid location. Therefore, we expect the equations derived for the various properties image of a top-hat jet at $T_p<T<T_{end}$ to be applicable also to all structured jets.

The motion of the centroid before the peak depends more strongly on the jet structure. The reason is that the point dominating the emission during the rising phase is outside the core, and its angle from the line of sight, $\xi$, increases with time in a way that depends on the specific structure. Yet, as we show below, the centroid location is close to $y_{max}$, and the location of $y_{max}$  can be bounded during this phase between $y_{max}$ of a spherical blast-wave and $y_{max}$ of a top-hat jet. To demonstrate that, we compare the value of $y_{max}$ of three different blast-waves: a top-hat jet and a structured jet, both with the same values of $\theta_{obs}$, $\theta_c$ and $E_{iso}(\theta<\theta_c)=E$, and a spherical blast-wave with energy $E$. We define $\xi_{sph}^{max}(T)$, $\xi_{sj}^{max}(T)$ and $\xi_{th}^{max}(T)$ as the angles from the line of sight to the region that emits the radiation observed at $y_{max}$, of the sphere, the structured jet and the top-hat jet, respectively. Since the energy profile of the structured jet decreases monotonically with the angle at $\theta>\theta_c$, there are two inequalities that are satisfied for any $T<T_p$:  $\xi_{sph}^{max}(T)\leq\xi_{sj}^{max}(T) \leq \xi_{th}^{max}(T)\approx \theta_{obs}-\theta_c$ and  $1 \approx \xi_{sph}^{max}\Gamma(\xi_{sph}^{max})<\xi_{sj}^{max}\Gamma(\xi_{sj}^{max})<\xi_{th}^{max}\Gamma(\xi_{th}^{max})$. These inequalities approach (roughly) an equality as $T\to T_p$. Now, from equations \eqref{eq:R_sky} and \eqref{eq:R},  we find that for any value of $T$, $\xi$ and $\Gamma(\xi,T)$: 
\begin{equation}\label{eq:y(xi,T)}
    y(\xi,T)=\frac{2 c T}{\xi}\frac{\Gamma^2\xi^2}{\frac{1}{1+2\omega}+\Gamma^2\xi^2} ~~.
\end{equation}
Note that as long as $\Gamma\xi\ge1$ the second term on the r.h.s. of this expression varies between 0.8 and 1 (for $\omega = \frac{3}{2}$), while the first term determines $y$. Thus, since for all three considered cases $\xi^{max}\Gamma(\xi^{max}) \geq 1$ the values of $\xi^{max}$ imply $y_{sph}^{max} \lesssim y_{sj}^{max} \lesssim y_{th}^{max}$ at any $T<T_p$ and all three become comparable at $T=T_p$. The final step is to note that similarly to a top-hat jet, during the rising phase of a structured jet, $y_{cen}\simeq y_{max}$, since the emission is dominated by a narrowly localized region within the structure near $\xi^{max}$ (we verify this numerically in appendix \ref{Appendix:Centroid_Edge}). Therefore we obtain $y_{sph}^{max} \lesssim y_{sj}^{cen} \lesssim y_{th}^{max}$.

The same line of reasoning can be followed to show that for a given $\theta_{obs}$ and $\theta_c$, a jet with a shallower structure (i.e., $E_{iso}$ that drops more slowly with $\theta$ outside of the core) will have at any given $T<T_p$ a higher value of $y_{max}$ (and $y_{cen}$), and therefore will deviate more from the solution of a top-hat jet. To demonstrate that, and to estimate the maximal possible deviation of the image centroid location of a given structured jet from that of a top-hat jet, we derive in appendix \ref{Appendix:power-law} an analytic solution for the value of $y_{max}$ of a jet with a power-law structure, $E_{iso}(\theta>\theta_c) \propto \theta^{-b}$. We find that in addition to the dependence on $\theta_{obs}-\theta_c$, these solutions depend on two more parameters, $b$ and $\theta_{obs}/\theta_{c}$. We also show in appendix \ref{Appendix:Centroid_Edge} that $y_{max}$ is an excellent approximation of $y_{cen}$.

\begin{figure}
    \centering
    \includegraphics[width = \columnwidth]{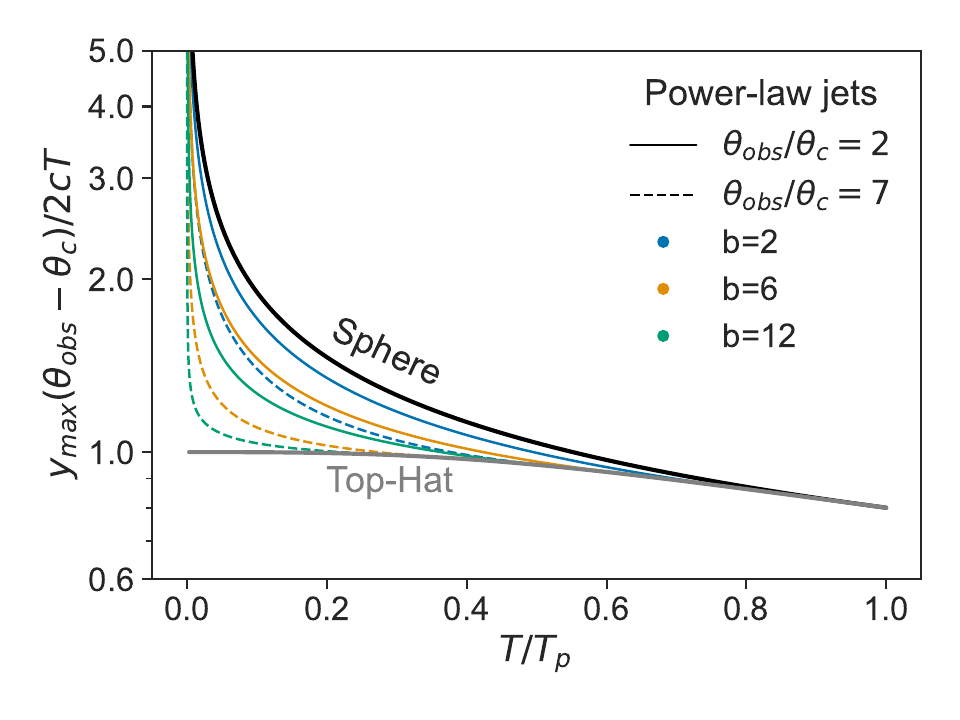}
    \caption{Normalized models for $y_{max}$ of a spherical blast-wave, a top-hat jet, and power-law jets with $E_{iso}\propto \theta^{-b},b=2,6,12$, for observers at two viewing angles. One can see that the steeper the power-law is, and the more the observer is off axis, the closer $y_{max}$ is to a that of a top-hat.}
    \label{fig:power-law}
\end{figure}

Fig. \ref{fig:power-law} depicts the value of $y_{max}$ of a power-law jet with several values of $b$ and two values of $\theta_{obs}/\theta_{c}$. This figure demonstrates several points. First, as expected, jets with shallower power-laws show a larger deviation from the solution of a top-hat jet, and they are all bounded by a spherical blast-wave. Second, The deviation from the top-hat solution is larger for smaller values of $\theta_{obs}/\theta_{c}$. Third, this figure can be used to estimate the error of using the top-hat approximation for a structured jet. For example, taking an extremely shallow structure with $b=2$, where most of the energy is outside the core, gives an estimated upper limit of the error. Fig. \ref{fig:power-law} shows that for a "barely" off-axis observer with $\theta_{obs}/\theta_j=2$, the maximal deviation at $0.2T_p$ is about 35\% while at $0.4T_p$ it is lower than 10\%. For an observer at $\theta_{obs}/\theta_j=7$ the maximal error is about 15\% at $0.2T_p$ and much lower than 10\% at $0.4T_p$. We therefore expect that for most purposes the approximation of the centroid location of  top-hat jets can be used for all structured jets starting at $\approx 0.2T_p$. For purposes that requires high accuracy one can use the formula given in appendix \ref{Appendix:power-law} to estimate the time this accuracy is achieved. Finally, it is important to remember that the error from the top-hat approximation is added to uncertainties and approximations used in the analytic derivation, which we explore below numerically. 

Before the time of the peak, the image of a structured jet consists of a curved bright region, which is brightest in the center, along the symmetry axis near $y_{max}$, and is fainter farther from that point. Both the curvature and the brightness distribution depend on the structure, making it difficult to model $\Delta x$ and $\Delta y$. However, as $T\to T_p$, the image must approach that of a top-hat jet seen after the time of the peak. We investigate this numerically in \S\ref{sec:Width} and \S\ref{sec:depth} and find that from $0.5T_p$ the top-hat model for $\Delta x,\Delta y$ offers a reasonable approximation for all jet structures.

\subsection{Applicability to other power-law segments}\label{sec: other PLS}
While the model was derived with $\nu_a,\nu_m<\nu_{obs}<\nu_c$ in mind, our model (or parts of it) can be applied to other parts of the spectrum as well. The power law segment can affect our model in two ways. First, the width of the ring seen in the image from a spherical blast wave depends on it, where the ring becomes narrower at steeper spectrum \citep{Granot1999,Granot1999b}. This will have a minor effect on the location of the centroid and the width of the image but can have a more significant effect on its depth. The second effect is that the power-law segment affects the shape of the peak, to the point that for some frequencies (e.g., $\nu_a<\nu_{obs}<\nu_m$ and $\nu_{obs}<\nu_a,\nu_m$)  the light curve does not peak when $\xi \approx \theta_{obs}-\theta_c$, where $\xi$ is the angle to the line of sight dominating the emission (i.e., the peak is not at the same time that it is observed at $\nu_a,\nu_m<\nu_{obs}$, which we denote here as $T_p$). Instead, at $T_p$ there is a break in the rise of the light curve  and the peak is seen at a later time (e.g., when $\nu_m$ crosses the observed frequency). Our conclusion is that our formulae for the location of the centroid and the image width are applicable, up to a correction factor of order unity, to all power-law segments given that the time at which $\xi \approx \theta_{obs}-\theta_c$ can be identified from the light curve. The formula for estimating $\theta_{obs}/\theta_c$ from the width of the peak (Eq. \ref{eq:Tend_T_p_ratio}) is applicable only for $\nu_m<\nu_{obs}$, where the peak is observed at $T_p$, with some corrections of the calibration coefficients for $\nu_m,\nu_c<\nu_{obs}$. At other segments it can be applied (with calibration correction) after modifying the definition of $T_{end}$ (Eq. \ref{eq:Tend}) so that $\frac{d\log F_{\nu}}{d\log T}$ is equal to the asymptotic power-law index of the light curve as seen by an on-axis observer at the same time.

\section{Numerical Simulations}\label{sec:Numerical_Simulations}

\subsection{Relativistic hydrodynamic simulations}
\subsubsection{Setup}
We used the publicly available code GAMMA\footnote{https://github.com/eliotayache/GAMMA} \citep{Ayache2022} to run 2D relativistic hydrodynamic (RHD) simulations. GAMMA uses an arbitrary Lagrangian-Eulerian approach in the main direction of the fluid motion, while keeping the grid static in the other direction. This approach makes GAMMA  extremely useful in the study of GRB jets, as the fluid motion in these jets is mainly in the radial direction. 
We used spherical coordinates, with a static polar grid and moving mesh with AMR in the radial direction. In this setup, we were able to resolve shocks with an initial matter Lorentz factor of 100. We used a setup similar to that described in in section 5 of \cite{Ayache2022}. See appendix \ref{Appendix:Numerical_Setup} for a detailed discussion of the numerical setup. 

To verify the grid resolution and AMR criterion used, we ran convergence tests. We also compared one of our simulations with a simulation on a static grid using the public Eulerian code PLUTO \citep{Mignone2007}, and found good agreement. See appendices \ref{Appendix:Convergance} and \ref{Appendix:PLUTO} for detailed discussion of the convergence tests and a comparison with PLUTO simulations.

\subsubsection{Initial conditions}
We simulated jets in a uniform medium, setting the energy, external density, and initial radius such that the entire jet structure has passed its deceleration radius, and the matter Lorentz factor right behind the shock in the core is 100 (corresponding to a shock Lorentz factor of $\Gamma_{core}\simeq 140$). We use an ideal gas equation of state with an adiabatic index $\hat{\gamma}=4/3$. The jet angular structures simulated include multiple power-law structures, top-hat, hollow and Gaussian jets (see structure definitions in table \ref{tab: sim_setup}). The power-law and top-hat simulations are simulated for an initial core angle of $\theta_{c,0}=0.05,0.15~\text{rad}$,  and the Gaussian has an initial core angle of $0.065$ rad. Given an energy angular structure $E_{iso}(\theta)$, we set the initial conditions at each angle as part of a Blandford-Mckee solution with the local value of $E_{iso}(\theta)$, in order to reduce the time it takes the jet to converge to a self-consistent structure. 

\subsection{Post-processing to Produce the light curves and Images}
We wrote a post-processing code that analyzes the results of the RHD code to produce images and light curves. In this code, we assume the standard afterglow model with an observing frequency above the synchrotron and self-absorption frequencies and below the cooling frequency $\nu_a,\nu_m<\nu_{obs}<\nu_c$. The code scans through every cell of the grid at every frame, and calculates its contribution to the luminosity and the image at any given observer time and observing angle. All simulations are post-processed for $p=2.05,2.2,2.5,2.8,3$, and the observer angles are selected as follows - simulations with $\theta_{c,0}=0.05$ and the Gaussian simulation are post-processed with $\theta_{obs}=0.1,0.12,0.15,0.2,0.3,0.4,0.5,0.6,0.7,0.85,1$, while simulations with $\theta_{c,0}=0.15$ are post-processed with $\theta_{obs}=0.3,0.36,0.45,0.6,0.75,1$. From these, only the simulations with a distinct peak and  $\theta_{obs}>2\theta_c$ (at the time of the peak) are selected. We present in the results only the range of observer times for which all the lab times that have a non-negligible contribution to the emission  at that observer time are within the scope of the simulation. This poses an additional constraint to the simulations used - simulations for which a non-negligible fraction of the emission at $T_p$ or $T_{end}$ is generated at radii that are not included in the RHD simulation are not used in our sample. The simulations used and the initial structures are listed in table \ref{tab: sim_setup}. 
The post-processing code is described in appendix \ref{Appendix:post_processing}. All figures presented in this paper are for $p=2.2$. Note that the light curve itself, and thus its calculation by the post-processing process, depends on the values of additional parameters, such as the source distance, the observing frequency, the jet total energy, the external medium density, and the fractions of energy in the magnetic field and in the electrons. However, all of our results depend on normalized light curves (i.e, the time is measured in units of $T_p$ and the flux by units of the peak flux) and these are independent of those parameters, as long as the observations are limited to a single power-law segment (e.g., \cite{Beniamini2020}).

\begin{table*}
    \centering
\begin{tabular}{|c|c|c|c|c|c|}
\hline 
Simulation & \multicolumn{2}{c|}{Initial $\frac{dE}{d\Omega}\propto$} & $\theta_{c,0}$ & $p$ & $\theta_{obs}$ range\tabularnewline
\hline 
\hline 
\multirow{5}{*}{Top-Hat} & \multicolumn{1}{c|}{\multirow{5}{*}{$\begin{cases}
1 & \theta\le\theta_{c,0}\\
0 & \theta>\theta_{c,0}
\end{cases}$}} & \multirow{5}{*}{} & \multirow{3}{*}{$0.05$} & $2.05-2.5$ & $0.1-1$\tabularnewline
 &  &  &  & $2.8$ & $0.12-1$\tabularnewline
 &  &  &  & $3$ & $0.15-1$\tabularnewline
\cline{4-6} \cline{5-6} \cline{6-6} 
 &  &  & \multirow{2}{*}{$0.15$} & $2.05$ & $0.3-0.6$\tabularnewline
 &  &  &  & $2.2-3$ & $0.3-0.75$\tabularnewline
\hline 
\multirow{12}{*}{Power-Law} & \multirow{12}{*}{$\begin{cases}
1 & \theta\le\theta_{c,0}\\
\left(\frac{\theta}{\theta_{c,0}}\right)^{-b} & \theta>\theta_{c,0}
\end{cases}$} & \multirow{2}{*}{$b=3$} & \multirow{2}{*}{$0.05$} & $2.05$ & $0.2-0.9$\tabularnewline
 &  &  &  & $2.2-3$ & $0.3-1$\tabularnewline
\cline{3-6} \cline{4-6} \cline{5-6} \cline{6-6} 
 &  & \multirow{6}{*}{$b=6$} & \multirow{3}{*}{$0.05$} & $2.05$ & $0.1-1$\tabularnewline
 &  &  &  & $2.2-2.5$ & $0.12-1$\tabularnewline
 &  &  &  & $2.8-3$ & $0.15-1$\tabularnewline
\cline{4-6} \cline{5-6} \cline{6-6} 
 &  &  & \multirow{3}{*}{0.15} & $2.05$ & $0.3-0.6$\tabularnewline
 &  &  &  & $2.2$ & $0.36-0.6$\tabularnewline
 &  &  &  & $2.5-3$ & $0.36-0.75$\tabularnewline
\cline{3-6} \cline{4-6} \cline{5-6} \cline{6-6} 
 &  & \multirow{4}{*}{$b=12$} & \multirow{3}{*}{$0.05$} & $2.05-2.5$ & $0.1-1$\tabularnewline
\cline{5-6} \cline{6-6} 
 &  &  &  & $2.8$ & $0.12-1$\tabularnewline
\cline{5-6} \cline{6-6} 
 &  &  &  & $3$ & $0.15-1$\tabularnewline
\cline{4-6} \cline{5-6} \cline{6-6} 
 &  &  & \multirow{1}{*}{$0.15$} & $2.05-3$ & 0.3-0.75\tabularnewline
\hline 
\multirow{2}{*}{Gaussian} & \multicolumn{1}{c|}{\multirow{2}{*}{$\exp\left(-\left(\frac{\theta}{0.07}\right)^{2}\right)$}} & \multirow{2}{*}{} & \multirow{2}{*}{$0.065$} & $2.05-2.5$ & $0.2-1$\tabularnewline
 &  &  &  & $2.8-3$ & $0.3-1$\tabularnewline
\hline 
\multirow{7}{*}{Hollow} & \multicolumn{1}{c|}{\multirow{4}{*}{$\propto\begin{cases}
\frac{1}{1+\exp\left(-5\left(\frac{\theta}{\theta_{c}}-\frac{1}{2}\right)\right)} & \theta\le\frac{\theta_{c,0}}{2}\\
1 & \frac{\theta_{c,0}}{2}<\theta\le\theta_{c}\\
0 & \theta>\theta_{c,0}
\end{cases}$}} & \multirow{4}{*}{} & \multirow{4}{*}{$0.05$} & $2.05$ & $0.1-1$\tabularnewline
 &  &  &  & $2.2$ & $0.12-1$\tabularnewline
 &  &  &  & $2.5-2.8$ & $0.15-1$\tabularnewline
 &  &  &  & $3$ & $0.2-1$\tabularnewline
\cline{2-6} \cline{3-6} \cline{4-6} \cline{5-6} \cline{6-6} 
 & \multirow{5}{*}{$\propto\begin{cases}
\frac{\theta}{\theta_{c,0}} & \theta\le\frac{\theta_{c,0}}{2}\\
1 & \frac{\theta_{c,0}}{2}<\theta\le\theta_{c}\\
\left(\frac{\theta}{\theta_{c,0}}\right)^{-b} & \theta>\theta_{c,0}
\end{cases}$} & \multirow{5}{*}{$b=6$} 
& \multirow{5}{*}{$0.15$} & &\tabularnewline
&  &  &  & $2.05$ & $0.3-0.6$ \tabularnewline
 &  &  &  & $2.2$ & $0.36-0.6$\tabularnewline
 &  &  &  & $2.5-3$ & $0.36-0.75$\tabularnewline
 &  &  &  & & \tabularnewline
\hline 
\end{tabular}
    \caption{Initial conditions and post-processing parameters for simulations used in this work. All simulations are post-processed for $p=2.05,2.2,2.5,2.8,3$. Simulations with $\theta_{c,0}=0.05$ and the Gaussian simulation are post-processed with $\theta_{obs}=0.1,0.12,0.15,0.2,0.3,0.4,0.5,0.6,0.7,0.8, 0.9,1$, while simulations with $\theta_{c,0}=0.15$ are post-processed with $\theta_{obs}=0.3,0.36,0.45,0.6,0.75,1$. The observing angles stated in the table are those for which there is an identifiable peak, and that both at $T_p$ and at $T_{end}$, all the expected emission is from within the simulated range of lab times.}
    \label{tab: sim_setup}
\end{table*}

\section{Numerical results}\label{sec:Numarical_Results}

We use the numerical simulations to test the applicability of our model to different jet structures, and to calibrate the constants defined in \S\ref{sec:Analytic}. In our analytic model, we use $\theta_c$ as the core angle of the jet without defining it properly. Moreover, in the simulations, since the jet structure evolves with time, so does the core angle and we need to find a general definition of $\theta_c$ that can be applied to all jet structures at all times. Therefore, we start by defining the core angle in \S\ref{sec:core_def}. In \S\ref{sec:Images} we review the simulated images, and then compare the analytical model to the numerical results for the centroid motion \S\ref{sec:Centroid}, image width \S\ref{sec:Width}, depth \S\ref{sec:depth}, and light curve \S\ref{sec:lightcurve}. 

\subsection{Defining the Jet Core Angle}\label{sec:core_def}
For a general jet structure the term "jet core" is not well defined since there is no definitive way to specify a location where the energy profile drops fast enough to be considered as the transition from the core to the "wings". We look for a physically meaningful definition that on the one hand can be applied to a general jet structure with any energy profile that drops monotonically (and fast enough) with the angle and on the other hand can be measured from the observations.
From theoretical point of view, considering a general jet energy structure, a natural distinction can be made between regions of the structure that are shallower than $E_{iso}\propto \theta^{-2}$ for which most of the energy is at large angles, and regions where the structure is steeper, for which the energy is dominated by small angles. Therefore a natural definition of the core angle is:
\begin{equation}\label{eq:core_def}
    \frac{d\log E_{iso}}{d\log \theta}\mid_{\theta=\theta_c}=-2.
\end{equation}
This definition promises that outside the core the energy is steep enough that most of the energy is in the core (or its immediate surroundings).

\begin{figure}
    \centering
    \includegraphics[width = \columnwidth]{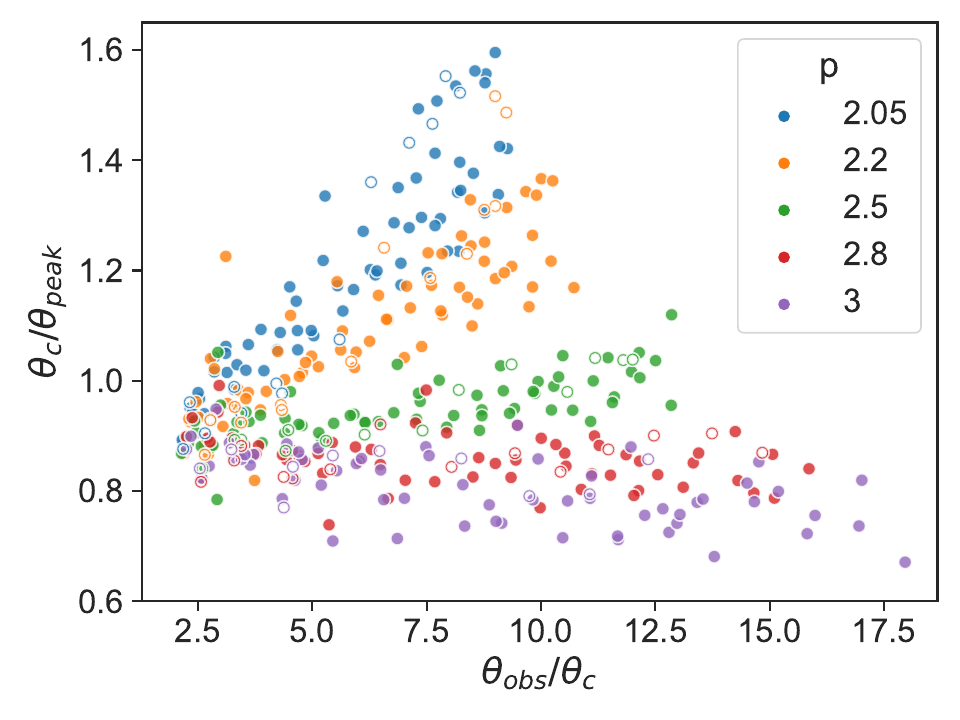}
    \caption{The core angle according to our definition (Eq. \ref{eq:core_def}), divided by $\theta_{peak}$ - the angle dominating the emission at the time of the peak. The hollow circles denote hollow jet simulations.}
    \label{fig:core_def}
\end{figure}

Considering observations, since one of the most identifiable observables is the peak of the light curve, and as theory predicts that at the time of the peak we observe (roughly) the edge of the jet core, we would like our definition to relate to the angle that dominates the emission at the time of the peak, $\theta_{peak}$. The exact criterion on the jet structure at $\theta_{peak}$ depends on $p$ and on $\frac{\theta_{obs}}{\theta_c}$ (see \citealt{Ryan2020}). However, it turns out that for relevant values of $p$ and $\frac{\theta_{obs}}{\theta_c}$, $\theta_c$ according to the definition of Eq. \ref{eq:core_def} provides a good approximation of $\theta_{peak}$. The quality of this approximation is shown in Fig. \ref{fig:core_def}. 

As the jet structure evolves with time, so does $\theta_c$ (as defined by Eq. \ref{eq:core_def}). We leave the study of this evolution to future work. Here we focus only on the value of $\theta_c$ at $T_p$, which is the only one that is accessible from the observations discussed in this paper. Therefore, unless stated otherwise, for each simulation and observing angle we use $\theta_c$ to denote the jet core angle at the time of the peak of the light curve and $\theta_{c,0}$ to denote the core angle at the beginning of the simulation\footnote{Note that in jets with steep wings (such as top-hat jets and steep power-law jets), the artificial structure causes the value of the core angle to drop initially on time scales that are shorter than the dynamical time, as energy flows sideways. The evolution of the core angle is reversed after the jet relaxes to a more stable structure that evolves on a dynamical time. This evolution causes $\theta_c<\theta_{c,0}$ in some of the simulations.}. In practice, we find $\theta_c$ of a given simulation by first selecting the lab time, $t$, dominating the emission at the time of the peak, and than integrating over the radial direction in the lab frame to find $E_{iso}(\theta)$ at that lab time, which we use to find the core angle according to Eq. \eqref{eq:core_def}.

\subsection{The Shape of the Image}\label{sec:Images}
\begin{figure*}
    \begin{center}
    \resizebox{.9\textwidth}{!}{%
    \includegraphics[height=3cm]{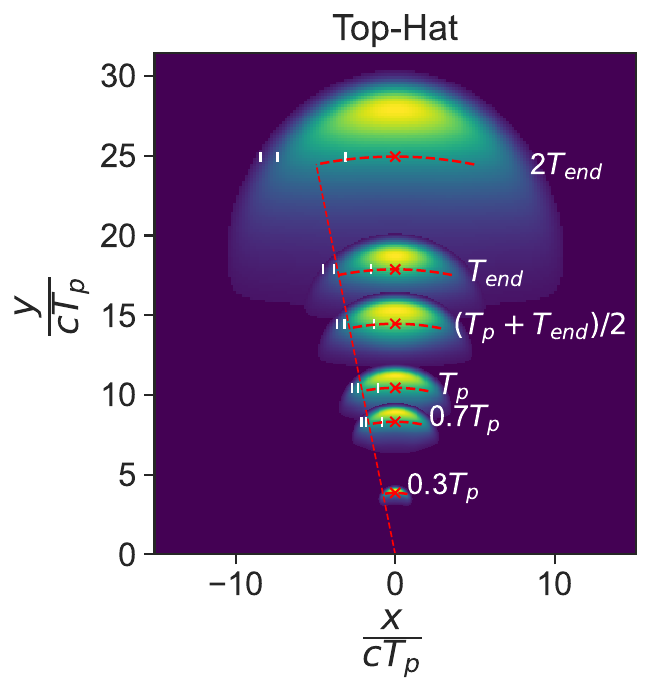}%
    \quad
    \includegraphics[height=3cm]{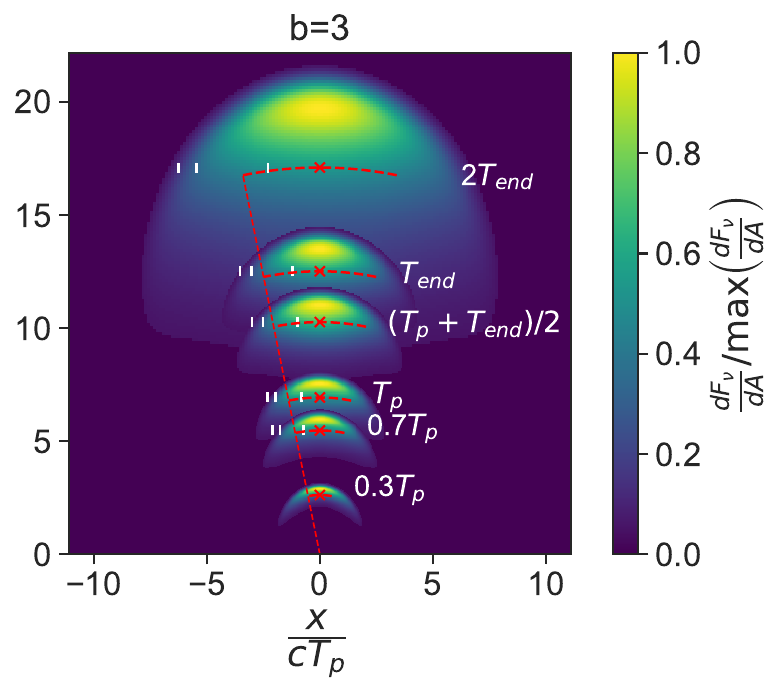}%
    }
    \end{center}    
\caption{Images of a top-hat jet (left) and a jet with a power-law structure of $b=3$ (right). Both jets have $\theta_{obs}/\theta_c \simeq 5$, so that the role of spreading is similar for both jets. The top-hat jet has a core angle (at the time of the peak) of $\theta_{c}=0.04$ rad and $\theta_{obs}=0.2$ rad, while the b=3 jet has a core angle (at the time of the peak) of $\theta_{c}=0.063$ rad and $\theta_{obs}=0.3$ rad. The flux in every image is scaled separately at each epoch. The white lines mark the width of the image within which $50\%,90\%$ and $95\%$ of the total flux is contained (integrated from the symmetry axis outwards).  The red $x's$ mark the flux centroid, the red dashed line marks an angle of $\theta_{c}/\theta_{obs}$ to the symmetry axis, and the red curves are arcs with the image centroid as the radius from the origin.}
    \label{fig:Image_evolution}
\end{figure*}

Before discussing specific properties of the the image, we examine its general shape. The analytic model predicts that the shape of the image during the peak phase ($T_p\le T\le T_{end}$) is an arc which is a part of a circle with a center at the origin, $(x,y)=(0,0)$, and a half opening angle of about $\theta_{c}/\theta_{obs}$. Fig.  \ref{fig:Image_evolution} shows the images of two extreme jets from our set of simulations - a top-hat jet, with no "wings", and a power-law jet with $b=3$, the most extended "wings". These figures show that the analytic prediction provides a fair description of the image, also at $T<T_p$. One source of deviation from the analytic description is sideways spreading of the jet. The spreading pushes matter to larger angles, and redistributes the jet energy so that material at larger angles propagates more slowly than the material along the axis. The light from this large-angle material causes the image to be dimmer farther from the axis and to have a slight crescent shape, instead of a pure arc. This deviation from the model is more prominent at late observer times and larger viewing angles. Another deviation from the analytic model is observed in the power-law jet (Fig. \ref{fig:Image_evolution}) where the light from the wings, is fainter and exhibits slower apparent motion than the core. This too causes the image to have a slight crescent shape, which is seen also at early times, before $T_p$.




\subsection{Centroid motion and measurement of $\theta_{obs}-\theta_c$}\label{sec:Centroid}
\begin{figure}
    \centering
    \includegraphics[width = \columnwidth]{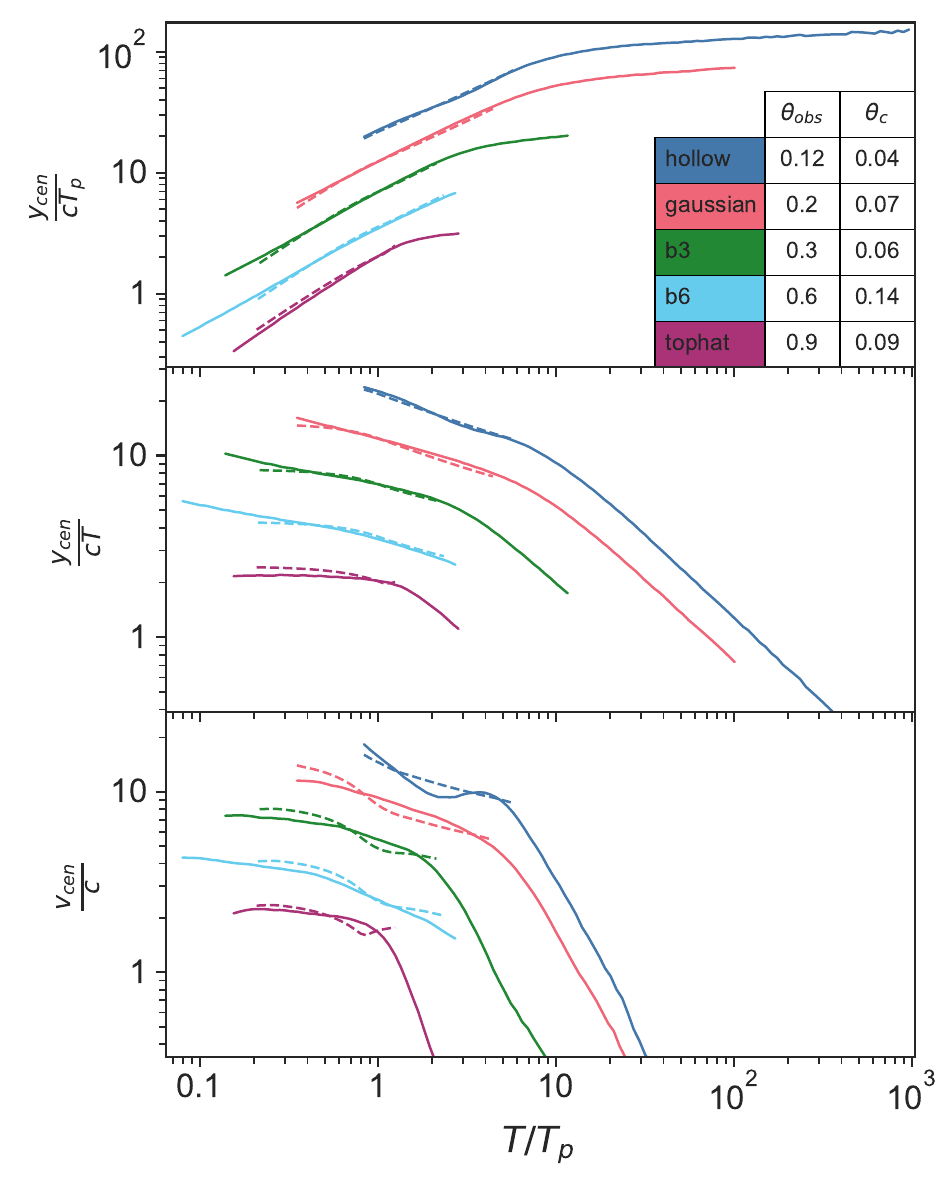}
    \caption{
    The image centroid location ({\it top panel}), average velocity ({\it middle panel}) and instantaneous velocity ({\it bottom panel})  as a function of the time (normalized by $T_p$). In all three panels, the simulations are plotted in solid lines and the calibrated models using $f_2$ (Eqs. \ref{eq:centroid_numerical} and \ref{eq:centroid_calib2}) in dashed lines. The model is plotted between $0.2T_p$ and $T_{end}$, and fits the simulations well. The centroid location and average velocity show deviation from the model by $\pm 10-15\%$ at most. Their maximal deviation is seen at $0.2 T_p$, when the effect of the jet structure is most significant. The instantaneous velocity is more strongly dependent on the structure, and deviates from the model more than the location and average velocity.
    }
    \label{fig:centroids_with_model}
\end{figure}

The analytic model of the centroid location can be written for $k=0$ (based on Eqs. \ref{eq:centroid_a} \& \ref{eq:centroid_b}) as:
\begin{equation}\label{eq:centroid_numerical}
y_{cen}=\frac{2cT}{\theta_{obs}-\theta_c}f\left(\frac{T}{T_p}\right) ~~~;~~~0.2T_p<T<T_{end}~,
\end{equation}
where $f(1)\approx 4/5$ and at $0.2T_p<T<T_p$ the function $f$ slowly decreases with time while at $T_p<T<T_{end}$ it drops more rapidly (roughly as $T^{-3/8}$). Shortly after $T_{end}$, the centroid motion is altered and can no longer be described by the model. This is because two of the main model assumptions break down. The emission zone stops traveling through the jet and becomes dominated by the region surrounding the jet axis, and more importantly, the region dominating the emission grows significantly as it includes a large region of the core (instead of an arc). The image becomes extended and the centroid is no longer close to $y_{max}$ (see Fig. \ref{fig:Image_evolution} and appendix \ref{Appendix:Centroid_Edge}). The result is that the centroid velocity drops much faster than the model predicts, and with the effect of spreading as the jets approaches non-relativistic velocities it even starts moving backward.

Fig. \ref{fig:centroids_with_model} shows the centroid location, average velocity and instantaneous velocity as a function of time for several simulations. All the image motions shown in the figure agree very well with the analytic predictions of Eq. \ref{eq:centroid_numerical} for  $0.2T_p<T<T_{end}$. 

\begin{figure}
    \centering
    \includegraphics[width = \columnwidth]{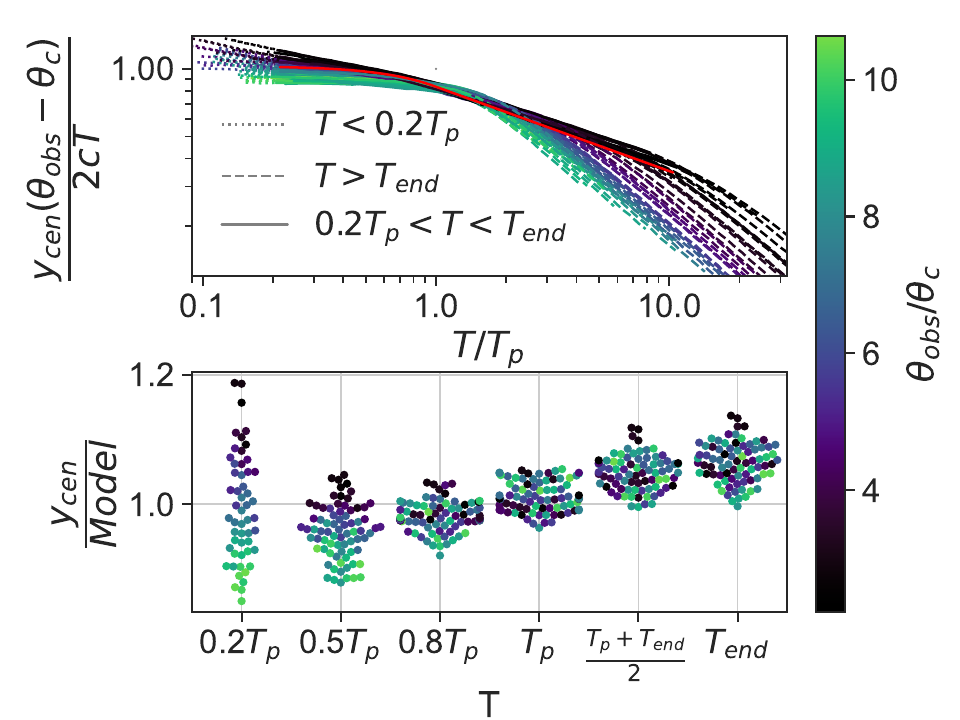}
    \caption{{\it Top panel}: the numerical value of $f$ (see Eq. \ref{eq:centroid_numerical}) is plotted for each of the simulations, alongside $f_1(\frac{T}{T_p})$ (Eq. \ref{eq:centroid_calib1}) in red. The period  $0.2T_p\le T\le T_{end}$ is marked with solid lines. Earlier and later times are marked with dashed lines. {\it Bottom panel}: the centroid location in all our simulations divided by the analytic model using $f=f_1$. Each dot at any time represents a single simulation. As expected, for $0.2T_p\le T\le T_{end}$, the simulations all roughly follow the same curve, and the model provides a good description for the simulations. At earlier times $T\le 0.2T_p$, the structure and $\theta_{obs}/\theta_{c}$ strongly affect the centroid location (as discussed in \S\ref{sec:structured_jets}). At $T>T_{end}$, the analytic model no longer holds and the simulations starts deviating more and more from the analytical curve. Note that while the model does not depend on $\theta_{obs}/\theta_c$, $T_{end}$ does, and therefore also the period in which the model is valid. }
    \label{fig:Centroid_f1}
\end{figure}

\begin{figure}
    \centering
    \includegraphics[width=\columnwidth]{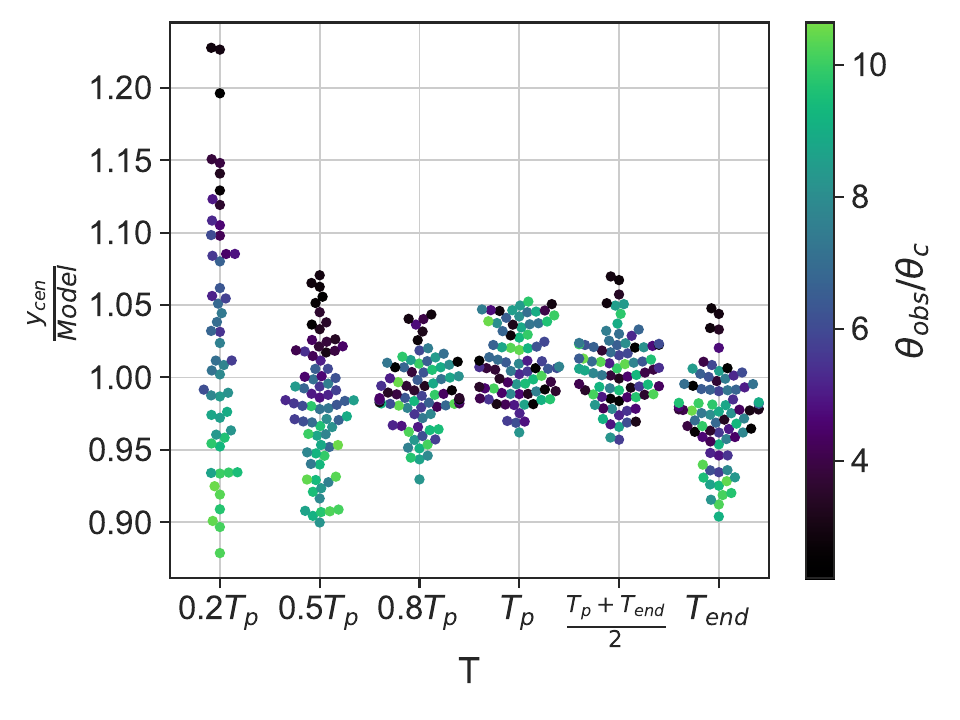}
    \caption{The centroid location divided by the analytic model (Eq. \ref{eq:centroid_numerical}) using $f=f_2$ (Eq. \ref{eq:centroid_calib2}). The marks are the same as in the bottom panel of Fig. \ref{fig:Centroid_f1}.}
    \label{fig:Centroid_f2}
\end{figure}

The numerical calibration of the analytic model is done by introducing calibration coefficients to the function $f$. Here we provide two models with different levels of calibration. The first is simpler and its accuracy is about 5\% at $T_p$ and better than 10\% at earlier and later times. The second calibration is slightly more complicated and its accuracy level is about 5\% at all times after $0.5T_p$. 
The simplest calibration of  Eqs. \ref{eq:centroid_a} \& \ref{eq:centroid_b} is by a single normalization factor:
\begin{equation}\label{eq:centroid_calib1}
f=f_1=C_{cen}\cdot\begin{cases}
\frac{1}{1+\left(\frac{T}{2T_{p}}\right)^{2}} & 0.2T_{p}\le T\le T_{p}\\
\frac{4}{5}\left(\frac{T}{T_{p}}\right)^{-\frac{3}{8}} & T_{p}<T\le T_{end}
\end{cases}
\end{equation}
Fig. \ref{fig:Centroid_f1} shows a comparison between the numerical results and the analytic model with $f=f_1$.
A more complex, but more accurate calibration is:
\begin{multline}\label{eq:centroid_calib2}
f=f_2\equiv\\
\equiv C_{norm}\cdot\begin{cases}
\frac{1}{1+\left(\frac{T}{2C_{T_{p}}T_{p}}\right)^{2}} & 0.2T_{p}\le T\le C_{T_{p}}T_{p}\\
\frac{4}{5}\left(1+C_{core}\frac{T-C_{T_{p}}T_{p}}{T_{end}-C_{T_{p}}T_{p}}\right)\cdot\left(\frac{T}{C_{T_{p}}T_{p}}\right)^{-\frac{3}{8}} & C_{T_{p}}T_{p}<T\le T_{end}.
\end{cases}    
\end{multline}
This calibration includes three constants, $C_{norm}$, $C_{core}$ and $C_{T_p}$, each one calibrates a different aspect of the analytic approximation. Fig. \ref{fig:Centroid_f2} shows a comparison between the numerical results and the analytic model with $f=f_2$. 

In both cases, the calibrations are performed with simulations with $\theta_{obs}\le 0.8$, and without the hollow jet simulations, since it seems that for $\theta_{obs}>0.8$ the model becomes significantly less successful in describing the simulations, and in some regions of the evolution, hollow jets can be dominated by emission from the inner walls of the jet and this can slightly alter the centroid motion. Yet, the model is reasonably accurate also for $\theta_{obs}>0.8$ and for the hollow jets. The values of all the calibration coefficients are listed in table \ref{tab:calibration}. 

\begin{figure*}
    \centering
    \includegraphics[width = \textwidth]{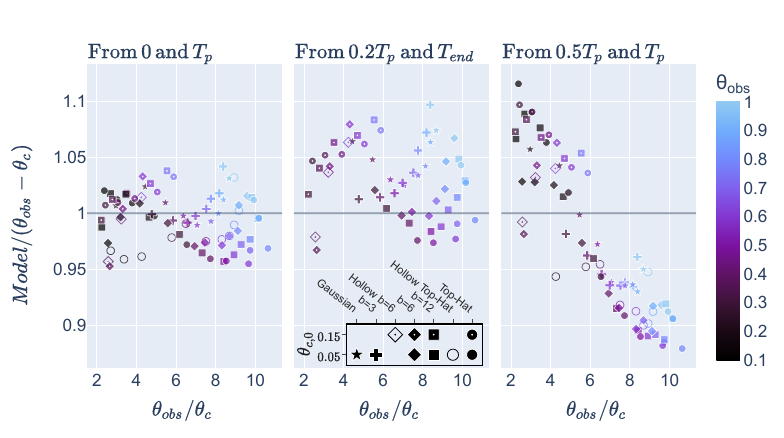}
    \caption{The value of $\theta_{obs}-\theta_c$ from the simulations is compared to the model for determining $\theta_{obs}-\theta_c$ using the centroid location at different choices of two epochs (Eq. \ref{eq:theta_obs-theta_c}).}
    \label{fig:obs_minus_core}
\end{figure*}

Our analytic model enables measuring $\theta_{obs}-\theta_c$ from measurements of the centroid location taken at two different epochs - $T_1,T_2$, where $T_1=0$ or $0.2T_p < T_1 < T_{end}$, and $0.2T_p < T_2 < T_{end}$. The quality of the measurement depends on the exact epochs that these measurements are taken. In order to estimate the effect of the epochs in which the two measurements are taken,  as well as the  systematic error that the uncertain jet structure induces in various observational scenarios we consider three possible pairs of epochs: $(T_1,T_2)=(0,T_p),~(0.2T_p, T_{end})$ and $(0.5T_p,T_p)$. To estimate the error we extract from the numerical images of each of the simulations the pair $y_{cen}(T_1)$ and $y_{cen}(T_2)$. We plug each pair into Eq. \ref{eq:centroid_numerical} with $f=f_2$ (Eq. \ref{eq:centroid_calib2}) and extract the analytic estimate of $\theta_{obs}-\theta_c$. We then compare this estimate to the actual value of  $\theta_{obs}-\theta_c$ in that simulation. Fig. \ref{fig:obs_minus_core} depicts the errors in the analytic estimate compared to the actual value of $\theta_{obs}-\theta_c$ for each of the simulations and for each of the three pairs of $T_1,~T_2$. It shows, first, that the model is most accurate for $(T_1,T_2)=(0,T_p)$, with an accuracy level that is better than about $\pm 5$\% for all jet structures and observer angles. For $(T_1,T_2)=(0.2T_p,T_{end})$  the accuracy of the analytic model is at the level of $^{+5\%}_{-10\%}$ and for  $(T_1,T_2)=(0.5T_p,T_{p})$ it
is at the level of $^{+15\%}_{-10\%}$. Interestingly, in the case of $(T_1,T_2)=(0.5T_p,T_{p})$, we see a strong correlation of the ratio $\theta_c/\theta_{obs}$ with the deviation of the analytic model. This implies that full modeling (e.g., using numerical simulations) of the centroid motion, that takes into account the value of this ratio (from the light curve) can be  accurate at a level of $\pm 5$\%. Finlay, the scatter of the numerical values for simulations with the same $\theta_{obs}/\theta_c$ but different jet structures provides an estimate of the systematic error that the unknown jet structure introduces. We see that in all cases this error is at a level of about $\pm 5$\% or better. 

\subsection{Image width}\label{sec:Width}

\begin{figure}
    \centering
    \includegraphics[width = \columnwidth]{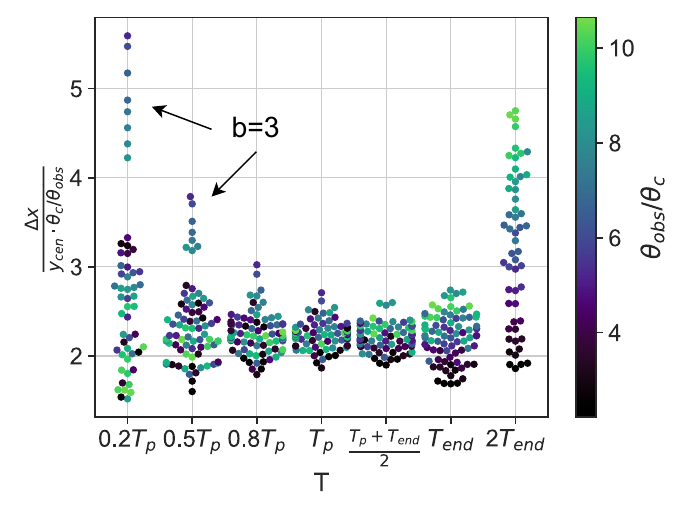}
    \caption{The image width from the simulations, defined as the symmetrical $\Delta x$ containing $90\%$ of the image flux, divided by the analytic model - Eq. \eqref{eq: width}. At early times, jets with a broader energy distribution have wider images, while at late times images with larger $\theta_{obs}/\theta_c$ are wider, as they have had longer time to spread.}
    \label{fig:width_over_centroid}
\end{figure}

We define the width of the image, $\Delta x = x_+-x_-$, as the symmetrical width ($x_+=-x_-$) containing $90\%$ of the total flux. Analytically, for $T_{p}\lesssim T\lesssim T_{end}$, $\Delta x\simeq 2\cdot y_{cen} \frac{\theta_c}{\theta_{obs}} $, and as seen in fig. \ref{fig:width_over_centroid} this  expression holds up to a factor of a 1.5 between $0.5 T_p$ and $T_{end}$. Before the time of the peak, the dispersion between simulations is mainly due to the jet structure, as jets with a shallower energy structure outside the core have a wider image. At later times, it can be seen that jets with larger $\theta_{obs}/\theta_c$ are wider, most probably due to lateral spreading not fully accounted for in the model. 
In appendix \ref{Appendix:depth} we present a calibrated model for $\frac{\Delta x}{y_{cen}}$.

As we show in \S\ref{sec:obs_core_graphs}, the image width can be used to measure $\frac{\theta_c}{\theta_{obs}}\cdot \frac{1}{\theta_{obs}-\theta_c}$, and can therefore be used alongside either the light curve or the image centroid motion (and a rough estimate of $T_p$) to measure $\theta_{obs}$ and $\theta_c$ independently.
\subsection{Image depth}\label{sec:depth}
\begin{figure}
    \centering
    \includegraphics[width = \columnwidth]{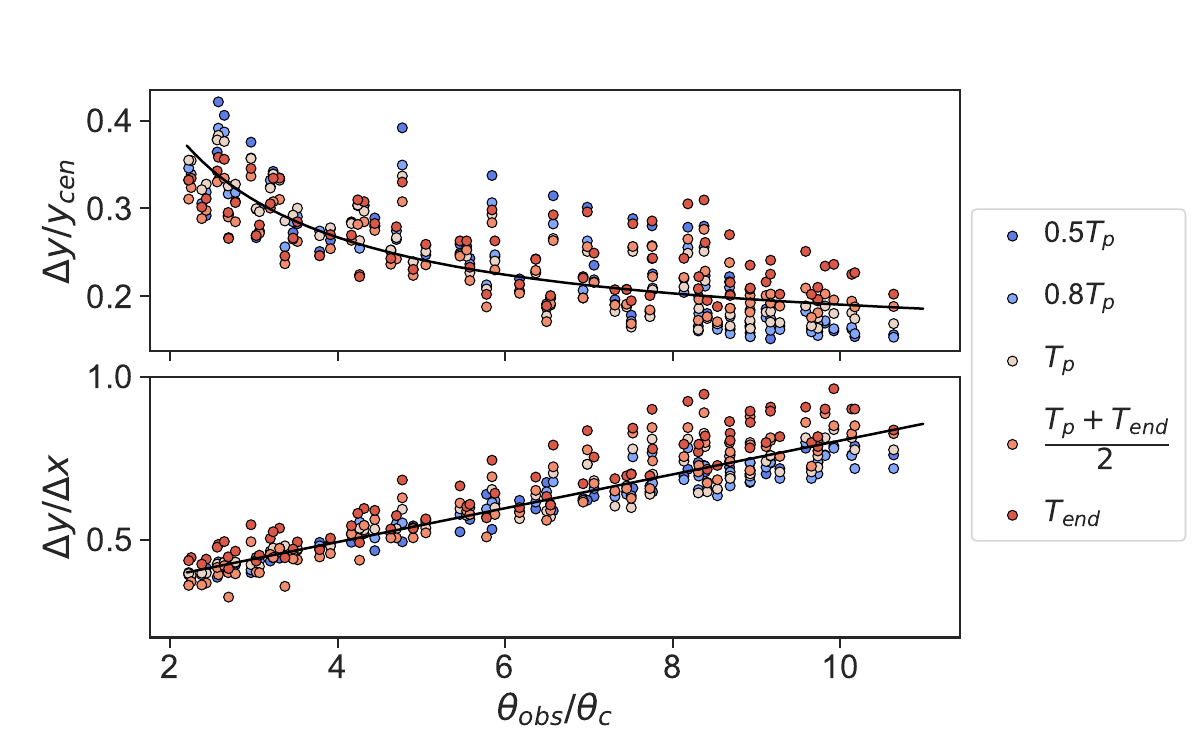}
    \caption{The depth to centroid location ratio (top panel) and the depth to width ratio (bottom panel) as a function of $\frac{\theta_{obs}}{\theta_c}$, in all our simulations. The black lines mark the calibrated relations described in appendix \ref{Appendix:depth}.}
    \label{fig:depth_over_width}
\end{figure}
We define the depth of the image as the smallest region $\Delta y$ containing 90\% of the flux. 
For $0.5 T_p \le T\le T_{end}$ and the range of $\frac{\theta_{obs}}{\theta_c}$ simulated, we find that $\frac{\Delta y}{\Delta x}$ can be reasonably approximated by a linear relation in $\frac{\theta_{obs}}{\theta_c}$, and accordingly, since $\Delta x\propto\frac{\theta_c}{\theta_{obs}} y_{cen}$,  $\frac{\Delta y}{y_{cen}}$ follows a linear relation in $\frac{\theta_{c}}{\theta_{obs}}$. A further discussion of these relations, and calibrated expressions appear in appendix \ref{Appendix:depth}.
In Fig. \ref{fig:depth_over_width}, these relations are plotted alongside the calibrated expressions.

\subsection{Finding $\theta_{obs}/\theta_c$ from the light curve}\label{sec:lightcurve}
\begin{figure}
    \centering
    \includegraphics[width = \columnwidth]{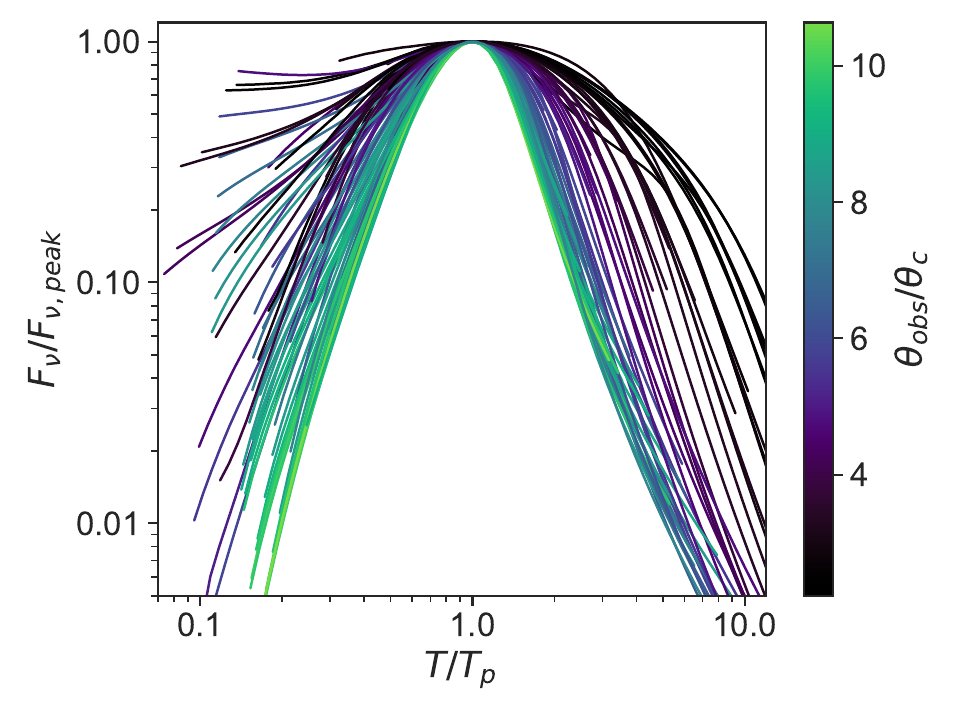}
    \caption{The normalized light curves of all the simulations are plotted, color coded by the value of $\theta_{obs}/\theta_c$. Before the time of the peak, when the structure dominates the evolution, and at late times, when the jets  become Newtonian, the light curve shapes are not correlated with $\theta_{obs}/\theta_c$. However,  the width of the peak is correlated with $\theta_{obs}/\theta_c$, as expected from Eq. \ref{eq:Tend_T_p_ratio}. This correlation saturates at $\theta_{obs}/\theta_c\simeq10$, as discussed in \S\ref{sec:lightcurve}.}
    \label{fig:lightcurves}
\end{figure}

As described in \S\ref{sec:angles_lightcurve}, $T_{p}$ and $T_{end}$ can be identified in the light curve and used to find the ratio $\theta_{obs}/\theta_c$. We use our simulations to calibrate the analytic expression that relates the light curve to this ratio.  In Fig. \ref{fig:lightcurves}, light curves for many different angle ratios and structures are plotted, and one can see the peak getting narrower as $\frac{\theta_{obs}}{\theta_{c}}$ grows. 

Asymptotically, when $\theta_{obs}/\theta_c\to \infty$, the duration of the phase in which the core dominates the emission approaches 0, and we would expect $T_{end}/T_p\to 1$. However, since $T_p$ is the time at which the light curve inclination vanishes and $T_{end}$ is found by $\frac{d\log F_{\nu}}{d\log T}=-p$, for a smooth light curve, the two times cannot coincide, and  $T_{end}/T_p> 1$. Therefore, for a large enough $\theta_{obs}/\theta_c$, the shape of the peak will only provide a lower limit on $\theta_{obs}/\theta_c$. Indeed, we find that for $T_{end}/T_p < 1.3$ ($\frac{\theta_{obs}}{\theta_c}\gtrsim10$) the shape of the light curve no longer depends on $\theta_{obs}/\theta_c$, and only a lower limit for $\theta_{obs}/\theta_c$ can be attained, which translates into an upper limit on $\theta_c$ and a good approximation of $\theta_{obs}\simeq\theta_{obs}-\theta_c$.

\begin{figure}
    \centering
    \includegraphics[width =\columnwidth]{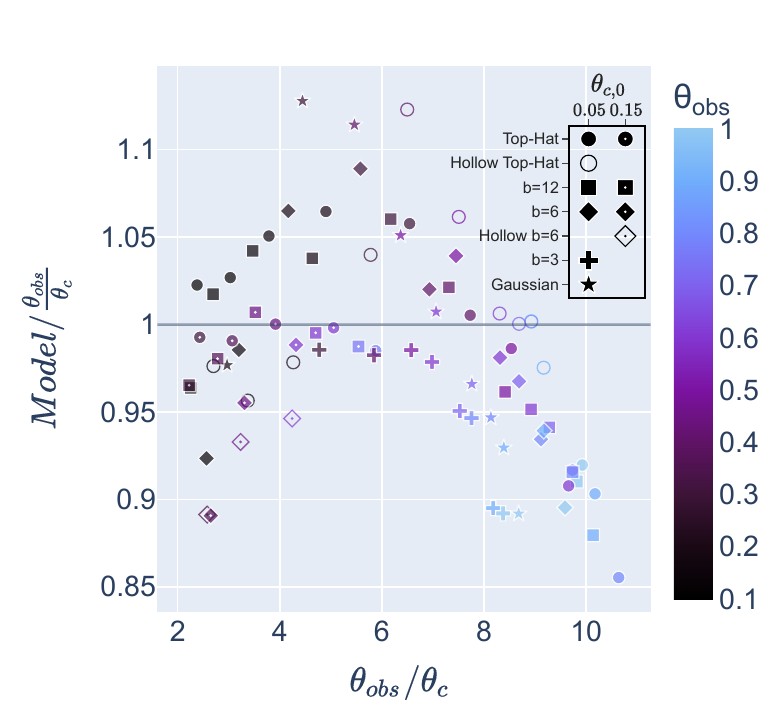}
    \caption{The value of $\theta_{obs}/\theta_c$ found from the light curve according to Eq. \eqref{eq:Tend} is compared to the values in the simulations.}
    \label{fig:theta_obs_o_theta_c}
\end{figure}

We calibrate the model using simulations with $\theta_{obs}\le 0.75$, and $\frac{T_{end}}{T_{p}}\ge 1.3$ (Equivalent to $\theta_{obs}/\theta_c\lesssim10$), and without simulations of hollow jets. The calibration constants are presented in table \ref{tab:calibration}. A comparison to simulations is presented in Fig. \ref{fig:theta_obs_o_theta_c}, which shows that when comparing the value of $\theta_{obs}/\theta_c$ inferred from $T_{end}/T_p$ to the simulation value, in most cases the error is $<10\%$.

\begin{table*}
    \centering

\begin{tabular}{|c|c|c|c|c|c|c|}
\hline 
\multirow{2}{*}{} & \multirow{2}{*}{Calibrtion Constant} & \multicolumn{5}{c|}{$p$}\tabularnewline
\cline{3-7} \cline{4-7} \cline{5-7} \cline{6-7} \cline{7-7} 
 &  & 2.05 & 2.2 & 2.5 & 2.8 & 3\tabularnewline
\hline 
\hline 
Centroid, minimal calibration (Eq. \ref{eq:centroid_calib1})& $C_{cen}$ & 1.01 & 1.03 & 1.07 & 1.09 & 1.11\tabularnewline
\hline 
\multirow{3}{*}{Centroid, full calibration (Eq. \ref{eq:centroid_calib2})} & $C_{T_{p}}$ & 1.03 & 1.1 & 1.33 & 1.57 & 1.74\tabularnewline
 & $C_{norm}$ & 0.99 & 0.99 & 0.99 & 0.98 & 0.98\tabularnewline
 & $C_{core}$ & 0.08 & 0.08 & 0.07 & 0.06 & 0.06\tabularnewline
\hline 
\multirow{2}{*}{Peak width - $\theta_{obs}/\theta_c$ (Eqs. \ref{eq:Tend_T_p_ratio})} & $C_{end}$ & 0.84 & 0.87 & 0.91 & 0.96 & 0.98\tabularnewline
 & $h$ & 0.39 & $\text{0.4}$ & $\text{0.4}$ & $\text{0.4}$ & $\text{0.4}$\tabularnewline
\hline 
\end{tabular}
    \caption{Calibration coefficients for analytic formulae of the centroid (Eqs. \ref{eq:centroid_calib1} and \ref{eq:centroid_calib2}), and for the relation between the light curve peak width and $\theta_{obs}/\theta_c$ (Eq. \ref{eq:Tend_T_p_ratio}).}
    \label{tab:calibration}
\end{table*}
\subsection{Determining $\theta_{obs}$ and $\theta_c$ independently}\label{sec:obs_core_graphs}
Using the light curve to determine  $\theta_{obs}/\theta_c$ (Eq. \ref{eq:Tend_T_p_ratio}) and the centroid motion to measure $\theta_{obs}-\theta_c$, we can solve for $\theta_{obs}$ and $\theta_c$ independently, and for each, find a model that depends only on $T_p,T_{end}$ and two measurements of the centroid location (see \S\ref{sec:summary of main results} for explicit expression of $\theta_{obs},\theta_c$ and $\theta_{obs}-\theta_c$).
\begin{figure*}
    \centering
    \includegraphics[width = \textwidth]{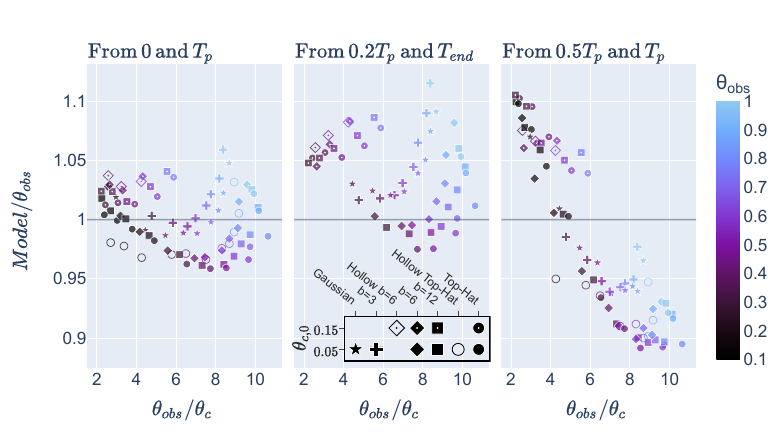}
    \caption{The observing angle found using the model (Eq. \ref{eq:theta_obs}) with the $f_2$ calibration, (Eq. \ref{eq:centroid_calib2}) divided by the value of $\theta_{obs}$. In each panel, the model uses the centroid displacement between the two times listed over the panel.}
    \label{fig:obs}
\end{figure*}
\begin{figure*}
    \centering
    \includegraphics[width=\textwidth]{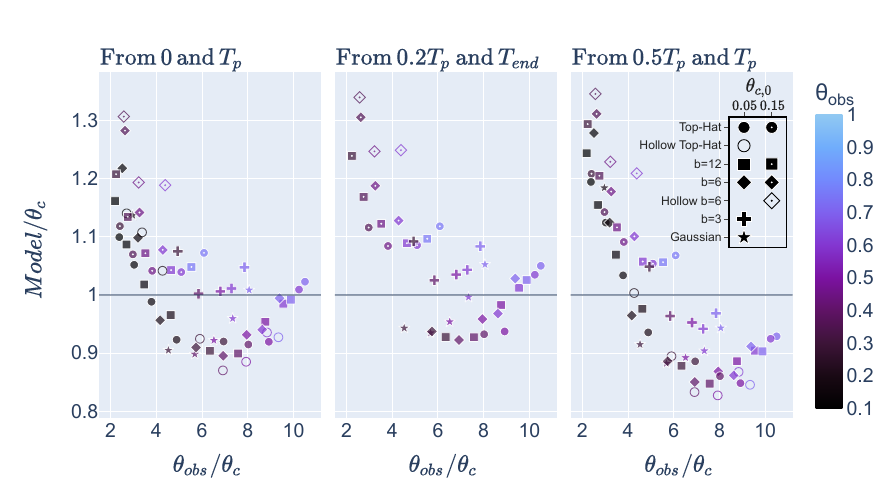}
    \caption{The jet core angle found using the model (Eq. \ref{eq:theta_c}) with the $f_2$ calibration, (Eq. \ref{eq:centroid_calib2}) divided by the value of $\theta_{c}$ according to the definition \eqref{eq:core_def}. In each panel, the model uses the centroid displacement between the two times listed over the panel.}
    \label{fig:core}
\end{figure*}
In Figs. \ref{fig:obs} and \ref{fig:core}, the model for $\theta_{obs}$ and $\theta_{c}$ is tested against the simulations, showing the errors one may expect when trying to constrain the two angles based on our analytic model. In these two figures, the angles are found using several different choices for the times $(T_1,T_2)$ between which the centroid displacement is measured. Using the centroid displacement between $0$ and $T_p$, or between $0.2T_{p}$ and $T_{end}$ the  error in $\theta_{obs}$ is $^{+5\%}_{-10\%}$ and in $\theta_c$ it is $^{+30\%}_{-10\%}$. Using the centroid displacement between $0.5T_p$ and $T_p$, the error in $\theta_{obs}$ is $\pm15\%$ and in $\theta_c$ it is $^{+30\%}_{-20\%}$. 

Figs. \ref{fig:obs} and \ref{fig:core} can also be used to estimate the systematic error in the measurement of $\theta_{obs}$ and $\theta_c$ due to poorly constrained jet structure. The idea is that the scatter of the errors of the comparison of the analytic model, which is independent of the jet structure, to the numerical simulations of jets with a large range of jet structures, provides an upper limit to this systematic error. Moreover, the error of the analytic model is often strongly correlated with the value of $\theta_{obs}/\theta_c$. This implies that a full numerical fit to the shape of the peak of a given light curve and to the centroid motion (taking into account various possible jet structures) would produce a scatter that is comparable to, or smaller than, that of the analytic model for a given value of $\theta_{obs}/\theta_c$. Thus, the scattering of the errors shown in Figs. \ref{fig:obs} and \ref{fig:core} for a given value of $\theta_{obs}/\theta_c$ gives an indication of the systematic error that a poorly constrained jet structure can introduce when a set of observations is modeled by full numerical simulations. For all the three pairs of $T_1,T_2$ the systematic errors in $\theta_{obs}$ are expected to be about $\pm 5\%$ and in $\theta_c$ they are about $\pm 15\%$.

\begin{figure}
    \centering
    \includegraphics[width=\columnwidth]{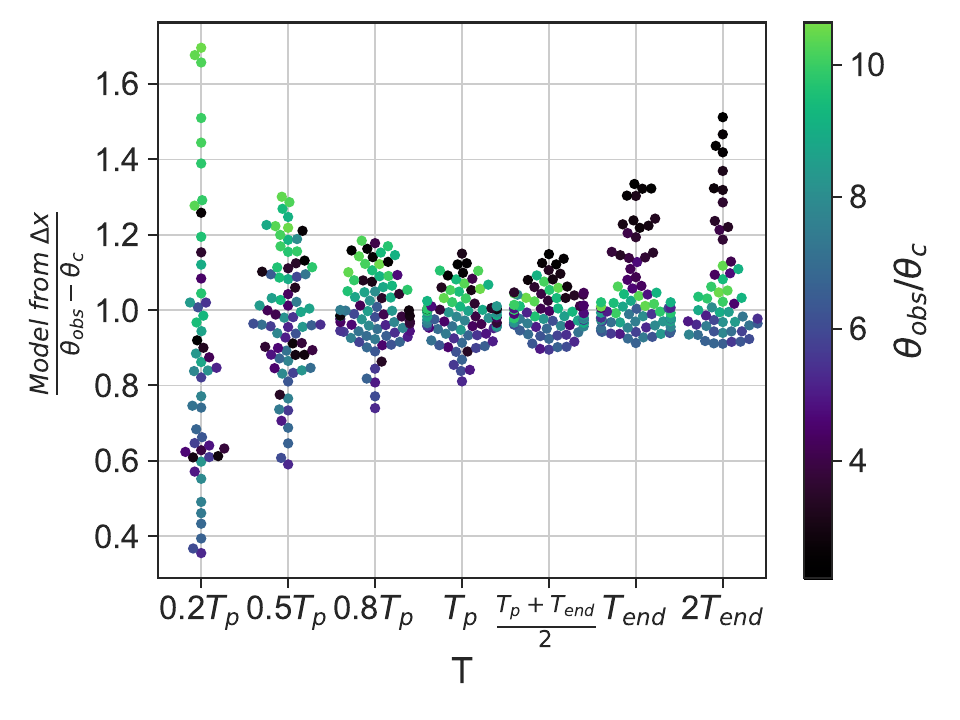}
    \caption{This graph demonstrates the ability to constrain $\theta_{obs}-\theta_c$ using the image width. $\theta_{obs}-\theta_c$ is found using the image width and light curve peak width in the following manner. The relation between the width and the centroid location (Eq. \ref{eq:width_calib}) alongside the calibrated expression for the centroid location (Eq. \ref{eq:centroid_numerical} with $f_2$ calibration \ref{eq:centroid_calib2}) give $\frac{1}{\theta_{obs}-\theta_c}\frac{\theta_c}{\theta_{obs}}$. $\theta_{obs}/\theta_c$ is found from the light curve peak width (Eq. \ref{eq:Tend_T_p_ratio}), and together, $\theta_{obs}-\theta_c$ can be determined. The graph portrays the ratio between $\theta_{obs}-\theta_c$ found as described above, and the simulation value at typical times.}
    \label{fig:width_obs_core}
\end{figure}
As described in \S\ref{sec:Image_theory}, the width of the image is proportional to the centroid location times $\frac{\theta_c}{\theta_{obs}}$. Therefore, a single measurement of the width alongside the light curve is sufficient to measure $\theta_{obs}$ and $\theta_c$. 
Fig. \ref{fig:width_obs_core} shows that without a specifically calibrated model (using the calibrated models for the centroid displacement and angle ratio, and the width-centroid relation calibrated by a normalization constant, as described in appendix \ref{Appendix:depth}) can recover $\theta_{obs}-\theta_c$ with a dispersion of $\pm 15-35\%$ due to the uncertainty in the jet structure. This method is most robust for jets with large ratios of $\frac{\theta_{obs}}{\theta_c}$, observed at $T\simeq T_p$.

\section{Using observation to determine $\theta_{obs}$ and $\theta_c$}\label{sec:core and obs}
\subsection{Summary of main results}\label{sec:summary of main results}
To determine $\theta_{obs}$ and $\theta_c$ using our analytic model, one needs the light curve observed at a frequency $\nu_a,\nu_m\le\nu\le\nu_c$ (See discussion in \S\ref{sec: other PLS} about using observations in other frequencies), a broad-band spectrum, in order to identify the electron power-law index, $p$, and the centroid displacement between two epochs (defined below). The calibration constants of our analytic formulae, which depend on $p$, are given in table \ref{tab:calibration}. 

The ratio $\theta_{obs}/\theta_c$ is deduced using Eq. \eqref{eq:Tend} by measuring the width of the light curve peak - the period during which the emission is dominated by the jet core, between when the flux is maximal, and when the observer starts seeing the asymptotic decline of the light curve. 

The flux centroid displacement between two times $T_1,T_2$ ($T_1=0$ and $0.2T_p\le T_2\le T_{end}$, or $0.2T_p\le T_1\le T_2\le T_{end}$ and $T_2\ge2 T_1$), provides a measurement of $\theta_{obs}-\theta_c$. Using Eq. \eqref{eq:centroid_numerical}, $\theta_{obs}-\theta_{c}$ can be written explicitly as:
\begin{equation}\label{eq:theta_obs-theta_c}
    \theta_{obs}-\theta_c = \frac{2 c}{y_{cen}(T_2)-y_{cen}(T_1)}\left(T_2 f\left(\frac{T_2}{T_p}\right) -T_1 f\left(\frac{T_1}{T_p}\right)\right) 
\end{equation}
Where $f$ is a dimensionless function, which may be replaced with $f_1$ (eq. \ref{eq:centroid_calib1}) for simple calibrations which depends only on $p$, or with $f_2$ (eq. \ref{eq:centroid_calib2}) for a more accurate calibration, which depends also on $T_{end}$. 
After obtaining $\theta_{obs}-\theta_c$, we use Eq. \eqref{eq:Tend_T_p_ratio} to find $\theta_{obs}$ and $\theta_c$:
\begin{equation}\label{eq:theta_obs}
\theta_{obs}=\left(\theta_{obs}-\theta_c\right) \left(\left(\frac{T_{end}}{C_{end}T_p}\right)^{1/h}+1\right)~,
\end{equation}
\begin{equation}\label{eq:theta_c}
\theta_{c}=\left(\theta_{obs}-\theta_c\right) \left(\left(\frac{T_{end}}{C_{end}T_p}\right)^{1/h}-1\right)~.
\end{equation}
Note that $\theta_c$ found in this manner is the core angle at the time of the peak, rather than at the time of jet injection. In our simulations, the same jet observed at a large angle may have a core angle 2 times larger than when observed at a small angle.

When considering observing strategies, one should consider the following points:

\begin{itemize}
    \item Observations of the centroid location at the origin and close to the time of the peak are least sensitive to the jet structure.
    \item If the origin of the centroid location  is not available, the two observations of the centroid location should be well separated, preferably by at least a factor of 2, as long as $T_2$ is not much larger than $T_{end}$.
    \item Between $0.2 T_p$ and $0.5 T_p$ the robustness of the model for the centroid location monotonously increases. 
    \item Light curve observations at high enough cadence are required around the peak and after it to properly identify $T_p$ and $T_{end}$.  
\end{itemize}

\subsection{Applying the model to the afterglow of GW170817}
The observations of the afterglow of GRB170817 consist of a rich light curve in radio, x-ray and optical, spanning from 9 to about 1000 days post-merger (\citealt{Makhathini2021} and references therein), a spectrum consisting of a single power-law, corresponding to $p=2.16\pm 0.01$ (1$\sigma$), VLBI observation of the centroid location at 75, 206 and 230 days post-merger \citep{Mooley2018,Ghirlanda2019}, and an HST observation of the centroid location of non-relativistic matter, 8 days after the merger \citep{Mooley2022}. 

Many studies have suggested values for $\theta_{obs},\theta_c$, either based on the light curve alone (which can only be used to measure $\theta_{obs}/\theta_c$, and not each value separately), or on the combination of the light curve and centroid motion. Most of these, used models that do not account for the jet spreading, with the exception of \cite{Hajela2019,Wu2019} who fit only the light curve using the boosted fireball model, \cite{Mooley2018} who use several PLUTO simulations to verify analytical models, and \cite{Mooley2022} who use a large sample of approximated hydro simulations (which include spreading). 

We apply our model to first find $\theta_{obs}/\theta_c$, then consider $\theta_{obs}-\theta_c$, and finally assess every angle separately. We use the calibration for $p=2.2$, since it is the closest value of $p$, but find that using the calibration of $2.05$ results in a correction much smaller than the error. 
To find $\theta_{obs}/\theta_c$ we must identify $T_p$ and $T_{end}$. We assume a spectrum with $p=2.16$ to normalize create a single light curve of the Chandra 1 keV x-ray observations \citep{Margutti2017,Troja2017,Troja2018,Troja2019,Troja2020,Nynka2018,Hajela2019,Haggard2017,Ruan2018,Piro2019} and the VLA 3 GHz radio observations \citep{Hallinan2017,Mooley2018b,Mooley2018c,Margutti2018,Dobie2018,Alexander2018,Makhathini2021}. Eye-balling the combined light curve, we can give a conservative estimate, $120~d \le T_{p}\le 162~d$ and\footnote{Note that if $T_{end}<230$ days then, formally, we cannot apply our model to the image centroid measurement at 230d. However, we do apply our model to this measurement as well since the error that this formal disagreement introduces is negligible.} $220~d\le T_{end}\le 266~d$. From these assessments, we find $5.4\le \theta_{obs}/\theta_c\le 11.2$. Note that the range of our limits on $T_p$ and $T_{end}$ is conservative. For example, \cite{Makhathini2021} estimate the time of the peak to be $T_{p}=155\pm 4~d$.

We can now apply our model to find $\theta_{obs}-\theta_c$ from the VLBI and HST observations of the afterglow centroid motion. There are four relevant observations. All quoted errors are $1\sigma$ unless specified otherwise. The first is an HST observation at 8 days, assumed to probe the origin location, since it observes emission from non-relativistic matter \citep{Mooley2022}. The others are VLBI observations at 75, 206 and 230 days, corresponding to 
$\simeq 0.5 T_p, 1.35 T_p,  T_{end}$, and displacements (relative to the HST observation) of $1.47\pm 0.32\times 10^{18} {\rm~cm}$, $2.49\pm 0.39\times 10^{18} {\rm~cm}$, $3.08\pm 0.44\times 10^{18} {\rm~cm}$ respectively, where the error includes statistical and systematic errors in the astrometric measurement of the centroid location, and the error in the distance to the host galaxy, taken here to be $40.7\pm 2.4$ Mpc \citep{Mooley2022}.

We fit our model to these three displacements using the least-squares method. Assuming $T_p$ and $T_{end}$ in the middle of the assessed range, $T_p=141~d$ and $T_{end}=243~d$, we find: $\theta_{obs}-\theta_c = 16.79 \pm 1.59\pm 0.84^\circ$  where the first error is from fitting the model to the observations (using $\chi^2$ test), while the second is a conservative estimate of the model systematic errors. We asses the systematic errors at 5\%, since for the relevant times and parameters, we find from Fig. \ref{fig:obs_minus_core} that our model gives rise to systematic errors of $2-5\%$ (with varying directions).
To asses the error due to the uncertainty in $T_{p},T_{end}$, we consider the extremal values that may be attained by selecting $T_{p},T_{end}$ within the range. For the minimal value of $T_p$ and maximal value of $T_{end}$ we find $\theta_{obs}-\theta_c = 15.81 \pm 1.5\pm 0.79^\circ$, while for the maximal value of $T_p$ and minimal value of $T_{end}$ we find $\theta_{obs}-\theta_c = 17.89 \pm 1.7\pm 0.89^\circ
$. We conclude that the error due to the uncertainty in $T_p$ and $T_{end}$ is smaller than $1^\circ$ and is therefore not the dominant source of error.  Taking all errors into account we obtain $\theta_{obs}-\theta_c = 16.8 \pm 2^\circ$. Altogether, our result is similar (including the range of errors) to that of \cite{Mooley2022}, who find that  $\theta_{obs}-\theta_c = 14-20^\circ$ (90\% confidence level).

Considering $\theta_{obs}$ and $\theta_c$ separately, we estimate the systematic errors due to the unknown jet structure from the scatter in Figs. \ref{fig:obs} \& \ref{fig:core}. For centroid location measurements at $T=0$ and at $T_p$ we estimate the error as $\pm 4\%$ for $\theta_{obs}$ and $\pm 10\%$ for $\theta_c$. To estimate the error due to the uncertainty in the values of $T_p$ and $T_{end}$, we find the extremal angle values that are obtained for possible values of $T_p$ and $T_{end}$. We find that using the minimal values of $T_p$ and $T_{end}$, $\theta_{obs}=18.97 \pm 1.8 \pm 0.76^\circ$ while when using maximal values of $T_p$ and $T_{end}$, $\theta_{obs}=19.8\pm 1.88 \pm 0.79 ^\circ$. This implies that error due to the uncertainty of $T_p$ and $T_{end}$ is subdominant (less that $0.5^\circ$), so we obtain
\begin{equation}
    \theta_{obs}= 19.4 \pm 2.1^\circ
\end{equation}
For $\theta_c$, using the lower limit for $T_p$ and upper limit for $T_{end}$, we find $\theta_c = 3.59 \pm 0.34 \pm 0.36^\circ$  and  using the upper limit for $T_p$ and lower limit for $T_{end}$, $\theta_c = 1.74\pm 0.17 \pm 0.17^\circ$. This implies that the uncertainty of $T_p$ and $T_{end}$ is the dominant error source, so we obtain
\begin{equation}
    \theta_c \simeq 1.5-4^\circ
\end{equation} 
Since our estimate of $T_p$ and $T_{end}$ is conservative we expect this estimate to be much better than $\pm 1\sigma$.
We stress that $\theta_c$ is the jet opening angle at the time of the peak and it is certainly possible that the jet opening angle upon launching was significantly smaller.

Our results are not in full agreement with those of \cite{Mooley2022}. They find that $\theta_{obs}=19-25^\circ$ and $\theta_c=4-6^\circ$ at 90\% confidence, while we find $\theta_{obs}=15.9-22.9^\circ$ at a 90\% confidence (assuming that the errors of $\theta_{obs}$ distributed normally) and $\theta_c \simeq 1.5-4^\circ$. As there is a full agreement in the estimate of $\theta_{obs}-\theta_c$, the source of this difference  is the estimate of the ratio $\theta_{obs}/\theta_c$ from the light curve.   \cite{Mooley2022} find $\theta_{obs}/\theta_c \simeq 4-5$ while we find that larger values are more likely. We see two possible reasons for this difference. The first is that \cite{Mooley2022} use approximated hydrodynamics, while we use a more accurate treatment of the fluid dynamics. The second, and more likely reason in our mind, is that \cite{Mooley2022} use a fit to the entire light curve to determine $\theta_{obs}/\theta_c$, while we use only the shape of the peak. A fit to the entire data gives weight to parts of the light curve that do not contain significant information on $\theta_{obs}/\theta_c$, such as the rising or the late decaying parts. In the case of GW 170817 the light curve at $T\gtrsim 300~d$ shows a decline of $F_\nu\sim T^{-1.92}$ \citep{Makhathini2021}, which is shallower than expected from the model. This deviation probably points to some deviation of the physics from the model (e.g., an additional energy source). However, the fitting procedure that minimizes the deviation of the model from all the data points, assumes that the model is correct at all times, even when it cannot fit the data well. Even though the light curve at late times is not affected by $\theta_{obs}/\theta_c$, the attempt to fit it (unsuccessfully) does affect the estimate of $\theta_{obs}/\theta_c$. 

The depth of the image of GW170817 may provide a meaningful lower limit on the value of $\theta_{obs}/\theta_c$ that is independent of the light curve. In Fig \ref{fig:depth_over_width}, one may see that if $\Delta y/y_{cen}\le 0.2$, that translates to a lower limit of $\theta_{obs}/\theta_c \ge 6$. \cite{Mooley2018} place an upper limit of $1$ mas on the image depth at 230 days post-merger. At this time, the centroid motion relative to the optical observation at 8 days is $5.07\pm 0.44$ mas, which corresponds to $\frac{\Delta y}{y_{cen}}\le 0.2$. No conclusive lower limit on $\theta_{obs}/\theta_c$ can be drawn from this since the definition for the depth upper limit used by \cite{Mooley2018} is not similar to our definition of the depths, however, this suggests that re-analysis of the observations using numerically predicted images will result in an independent lower limit on $\theta_{obs}/\theta_c$.

\section{Systematic errors in constraints on the Hubble constant} \label{sec:H0}
In light of the tension between measurements of the Hubble constant by distance ladders in the local universe \citep{Riess2022,Freedman2019} and by the CMB \citep{Planck2020}, there is growing interest in using GW events to measure $H_0$. The main advantage of these sources is the usage of the GW signal, rather than a distance ladder, to measure distances in the local universe \citep{Schutz1986}. One of the main sources of error in this measurement is a degeneracy between the distance and the inclination of the binary orbital plane with respect to the line of sight. 
For events at a small inclination ($i \lesssim 50^\circ$), this degeneracy takes the form (to a high degree of accuracy, regardless of the GW polarization)  
\begin{equation}
    h\propto\frac{\cos i}{d_L}
\end{equation}
where $h$ is the strain, $i$ is the binary inclination, and $d_L$ is the luminosity distance to the GW source. 

One way to remove this degeneracy is by measuring the angle between the jet symmetry axis and the line of sight, $\theta_{obs}$. Then, under the assumption that the jets are aligned with the orbital angular momentum, we obtain $\cos(i)=\cos(\theta_{obs})$ \citep{Hotokezaka2019,Mastrogiovanni2021,Wang2022,Bulla2022}. This method was used by \cite{Hotokezaka2019} to reduce the errors on the measurement of $H_0$ from the observations of GW170817 from about 15\% \citep{LIGO_H02017} to 7\%.
There are many challenges and uncertainties in measuring $H_0$ using this method to an accuracy that can lift the Hubble tension. These include, just to list a few, a low event rate, insufficient data quality, possible misalignment between the jet and the orbital plane and various observational biases. However, even if we overcome all these challenges there is a concern that the systematic uncertainty due to the unknown jet structure would be too large, making it impossible to obtain a 2\% error on $H_0$ using this method. Our study enables us to estimate the systematic error induced by the unknown jet structure on the measurement of $\cos(\theta_{obs})$ and thus on $H_0$.

\begin{figure*}
    \centering
    \includegraphics[width = \textwidth]{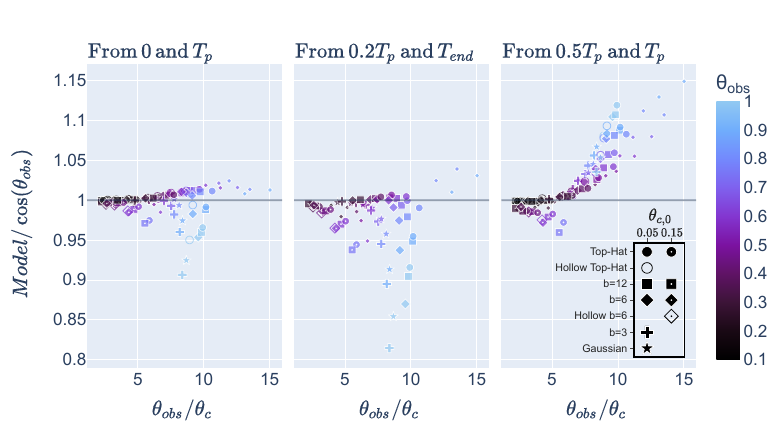}

    \caption{The cosine of the observer angle found by applying the model (Eq. \ref{eq:theta_obs}) and using the centroid location at the times listed above the panel, divided by the value of $\cos (\theta_{obs})$. The small diamonds denote $b=6$ simulations with non-constant microphysical parameters (As discussed in \S\ref{sec:micro_phys})}
    \label{fig:cos_obs}
\end{figure*}

Fig. \ref{fig:cos_obs} shows the ratio between the value of $\cos(\theta_{obs})$ measured by our model (which is ignorant to the jet structure) and its actual value in our simulations. The scatter of this ratio, for a given value of the ratio $\theta_{obs}/\theta_c$ provides an estimate of the systematic error. From this figure, we find that for $\theta_{obs}\le 0.7$ rad the systematic uncertainty in $\cos(\theta_{obs})$ is better than $\pm 1.5$\%  when the centroid displacement is measured between 0 and $T_p$ and about $\pm 2-3\%$ when the centroid location is not measured at $T=0$. For $0.7 \le \theta_{obs}\le 1$ rad the error can be much larger (as high as $\pm 5\%$). We conclude, that for jets that are not too far off axis the systematic uncertainty due to the jet structure can be controlled, and is probably not expected to be the leading source of systematic error in the measurement of $H_0$ by this method. 

\section{The effect of non-constant microphysical parameters}\label{sec:micro_phys}
In our model, we assumed that the microphysical parameters are constant. So far, there is no clear observational signature of evolution of the microphysics  in  GRB afterglow. However, there is no systematic study that tests how much evolution of the microphysical parameters is consistent with the available data. Therefore, it is possible that there is, a so far undetected, evolution. Moreover, a comparison of GRB afterglows and supernova radio counterparts suggests that there are differences in the microphysical parameters between relativistic and Newtonian shocks (at least in the value of $p$), suggesting that at least while the shock is mildly relativistic  some of the microphysical parameters vary with the shock velocity. 

In this section we examine the effect of unaccounted for evolution of the microphysical parameters on the estimates of $\theta_{obs}$ and $\theta_c$, and especially by how much it affects the systematic error of $\cos(\theta_{obs})$. Observationally, an evolution of $p$ is relatively simple to detect via the observed spectrum. For example, in the afterglow of GW170817 $p$ seems to be constant to a very high accuracy during the entire observed evolution. However, an evolution of $\epsilon_e$ and $\epsilon_B$ is much harder to detect. The reason is that the signature of such evolution can be seen only after the peak of the light curve, since before the peak any observation can be explained by the unknown jet structure. However, also after the peak other effects may be responsible for deviation of the light curve from the model. Therefore, here we focus on evolution of these two parameters.  Since the most likely parameter that evolves with time and may affect the value of the microphysical parameters is the shock Lorentz factor, we consider four cases, varying $\epsilon_{e,B}$ separately in either an increasing or a decreasing power of $\gamma\beta$: $\epsilon_{e,B}\propto \gamma\beta,\left(\gamma\beta\right)^{-1}$. For $p=2.2$, this corresponds to the light curve having an asymptotic power-law decline with an index of $1.8$ and $2.6$ accordingly if varying $\epsilon_B$ or of $1.6$ and $2.8$ if varying $\epsilon_e$. Such deviations in the decline rate of the decaying phase are easily detectable. We test the effect of evolving parameters on a $b=6$ power-law jet with $\theta_{c,0}=0.05$.

Examining the results of the simulations we find that the main effect of evolution of the microphysical parameters is that the time of the peak is altered, and with it, the value of the core angle at the peak  (due to spreading). For example, for a viewing angle of $\theta_{obs}=0.2$ rad, the peak time is shifted by a factor of 0.7-1.4 and the core angle by a factor of 0.8-1.15 compared to a similar jet with constant microphysical parameters. At larger observing angles the effect on the peak time decreases, so that the peak time is altered by $\pm 20\%$ for $\theta_{obs}=0.8$ rad, and the value of  $\theta_c$ is revised by a factor of 0.7-1.6 of its value for the constant microphysical parameters. 

\begin{figure}
    \centering
    \includegraphics[width = \columnwidth]{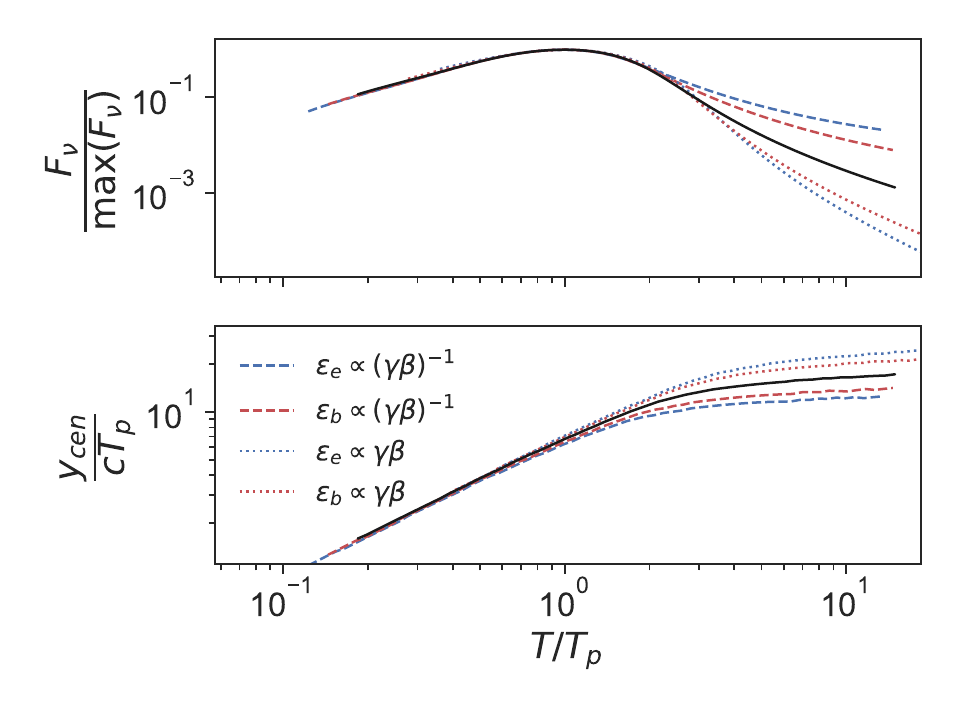}
    \caption{The normalized light curve (top panel) and centroid motion (bottom panel) for a $b=6$ power-law jet, observed at $\theta_{obs}=0.3$. each curve corresponds to a different dependence of the microphysical parameters on $\gamma\beta$, as listed in the legend. The black curves correspond to constant microphysical parameters. The effect of non-constant microphysics starts becoming noticeable at $T_p$, and it becomes significant only after $T_{end}$}
    \label{fig:micro_phys}
\end{figure}

The estimate of $\theta_{obs}$, however, depends on $y_{cen}(T\sim T_p)$ and on the normalized light curve (time measured in units of $T_p$). Fig. \ref{fig:micro_phys} depicts the normalized light curves and centroid displacements from simulations with evolving $\epsilon_{e,B}$, plotted for an observing angle of $0.3$ rad. This figure shows that until $T_{end}$ the normalized light curve is barely influenced by the changes in $\epsilon_e,\epsilon_B$. After $T_{end}$ the effect of the varying microphysical parameters is evident, and the asymptotic decline of the light curve is altered as expected. The centroid motion is not effected much until $T_p$. By $T_{end}$ the centroid location is altered by up to $\pm 10\%$, and a more significant effect is seen after $T_{end}$. The effect on $T_p/T_{end}$ is minor because this ratio is only sensitive to a short time relative to the systems dynamical time. The centroid location is robust as it only probes the difference between the integrated motion of two points, each of which moved at an apparent velocity of $ \frac{2}{\xi}$ (where $\xi$ is the angle of the point to the line of sight) for most of their evolution, and is only effected by the parameters' evolution for a short time.

To evaluate the effect of unaccounted for evolution of $\epsilon_{e,B}$ we apply our analytic formula (calibrated using simulations with constant microphysics) to estimate  $\theta_{obs},\theta_c$ and $\cos (\theta_{obs})$ and compare it to the actual values in the simulations.  We find that for most of the simulations the error of the analytic formulae due to the evolving microphysics are comparable to, or at most only slightly larger than, the errors found in simulations with constant microphysics due to the different jet structures (shown in Figs. \ref{fig:obs}, \ref{fig:core} and \ref{fig:cos_obs}). This can be seen for the estimates of $\cos (\theta_{obs})$ in Fig. \ref{fig:cos_obs}, which includes also the errors of simulations with evolving $\epsilon_{e,B}$.  

These errors can be viewed as upper limits on systematic errors arising from non-constant microphysics parameters in systems with light curves which behave roughly as expected for constant parameters. Note that a difference (of unknown origin) between  the asymptotic decline of the afterglow of GW170817 and the model prediction was detected, despite being much smaller than the cases considered here. From this we can conclude that variation in the microphysical parameters that does not alter the light curve significantly most probably does not cause a significant systematic error in constraining $\theta_{obs}$, $\theta_c$, and most importantly $\cos (\theta_{obs})$. 

\section{Conclusions}\label{sec:Conclusions}
In this work we study the afterglow images of off-axis GRBs, and their relation to the jet geometry. We present three main results. The first, a detailed study of the images of off-axis jets, which we use to show that the jet core angle and observing angle can be measured using any two of the three observables: the light curve around the peak, the flux-centroid motion and the image width. The second, a numerically-calibrated analytic model for finding the jet core angle and viewing angle of off-axis jets, using the afterglow light curve and flux-centroid motion (this method is summarized in \S\ref{sec:core and obs}). And the third, the systematic errors expected in $\theta_{obs}$, $\theta_c$ and $\cos(\theta_{obs})$ due to uncertainty in the jet angular structure, which we determine by comparing our model to a large sample of 2D relativistic hydrodynamic simulations with diverse jet angular structures.

Our calibrated formulae are restricted to observations at frequencies $\nu_a,\nu_m\le\nu\le\nu_c$, as expected for off-axis radio, and possibly also optical and X-ray, afterglows. However, the analytic formulae provide useful approximations also for frequencies at other power-law segments (see discussion in \S\ref{sec: other PLS}). To derive quantitative formulae of $\theta_c$ we needed first to provide a general, physically motivated, definition of $\theta_c$, which is given in Eq. \ref{eq:core_def}. Anywhere in this paper, unless stated otherwise, $\theta_c$ refers to the core angle according to this definition at the lab time which dominates the emission at the time of the peak of the light curve $(T_p)$. Note that this value is larger than the jet core angle as it starts propagating in the circum-merger medium after it expands following its breakout from the merger ejecta. 
Below, we discuss our main results on each of the topics we studied. \\

{\bf Image properties}:  The image of an off-axis jet can be described (approximately) during most of the evolution (at least between $0.2T_p$ and $T_{end}$) as a bright arc with the following properties:
 \begin{itemize}
    \item The centroid location (which is close to the arc radius) is approximately $y_{cen} \sim \frac{2cT}{\theta_{obs}-\theta_c}$. This approximation is accurate up to a factor $\lesssim 2$ that depends mostly on $T/T_p$ and is given at two levels of accuracy in Eqs. \ref{eq:centroid_calib1} and \ref{eq:centroid_calib2}.   
    \item The width of the image is smaller than the centriod location by a factor that proportional to $\frac{\theta_c}{\theta_{obs}}$. More accurately $\Delta x \approx y_{cen}\frac{2\theta_c}{\theta_{obs}}$. this approximation is most accurate near the time of the peak.
    \item The depth depends on $\theta_{obs}/\theta_{c}$ while $\frac{\theta_{obs}}{\theta_c}$ is small enough, while for large values of $\theta_{obs}/\theta_c$, the depth provides only a lower limit on the angle ratio. 
 \end{itemize}

{\bf Measuring $\theta_{obs}$ and $\theta_c$:} In this paper we provide calibrated, structure independent, formulae for three observables, each of which constrain a different combination of $\theta_{obs}$ and $\theta_c$. The width of the peak of the light curve measures  $\theta_{obs}/\theta_c$, the location of the image centroid measures $\theta_{obs}-\theta_c$ and the width of the image measures $(\theta_{obs}-\theta_c)\cdot \frac{\theta_{obs}}{\theta_c}$. Thus, any two out of these three observables are enough to measure $\theta_{obs}$ and $\theta_c$\footnote{Using the centroid motion and width requires a rough estimate of the time of $T_p$ and possibly also of $T_{end}$, however, for this sake the light curve does not need to be sampled at high cadence.}. 
To obtain each of these observables the following observations are needed: \textit{The width of the peak} requires identification of two times in the light curve, the peak time, $T_p$ and the time when the light curve starts its asymptotic decline, $T_{end}$. Note that since the asymptotic decline depends also on $p$, this observation also requires a broad-band spectrum at around the time of $T_{end}$.  \textit{The image centroid displacement} between two times, either $T_1=0$ and $0.2 T_p\le T_2\le T_{end}$ or $0.2 T_p\le T_1 < T_2\le T_{end}$ and $T_2 \gtrsim 2 T_1$ provides a measurement of $\theta_{obs}-\theta_c$. Finally, \textit{the image width} reasonably constrains the geometry if it is measured at $0.5T_p\le T\le T_{end}$, and most robustly near the time of the peak. Comparison of our analytic formulae to numerical simulations, shows that based on observations of the light curve and the centroid motion as described above, our formulae for $\theta_{obs}$ and $\theta_c$ can be accurate to a level of $5-10\%$ and $30\%$, respectively. As our formulae are independent of the jet angular structure, our errors includes a complete ignorance of this structure.

{\bf Systematic Errors}: We use our results to study the systematic errors in measuring $\theta_{obs}$, $\theta_c$ and most importantly $\cos (\theta_{obs})$, due to the uncertainty in the jet structure. These errors are most interesting in the context of using constraints on $\cos (\theta_{obs})$ to reduce the errors in measurements of $H_0$ by GW-EM emission of binary mergers. The estimate of the systematic errors is done by applying our analytic formulae to our simulations, and examining the scatter of the errors for a given value of $\theta_{obs}/\theta_c$. Our analytic formulae are agnostic of the jet structure. Therefore, the scatter of the errors found by testing the  analytic predictions against numerical results of simulations with a large range of jet structures, provides an upper limit on the systematic error that the uncertainty in the jet structure can introduce when one carries a complete numerical fit to the entire data set. We find that for $\theta_{obs}\le 0.7~\text{rad}$ the systematic uncertainty in $\cos(\theta_{obs})$ is better than $\pm 1.5$\%  when the centroid displacement is measured between 0 and $T_p$ and about $\pm 2-3\%$ when the centroid location is not measured at $T=0$. This means that the systematic error due to the uncertain jet structure is low enough to obtain an accurate measure of $H_0$ from a sample of GW events with jets that are not pointing too far away from us. 

To address another possible source of systematic errors we study the effect of temporal evolution of the microphysical parameters. We investigate a case where $\epsilon_e$ or $\epsilon_B$ depend on the shock proper velocity in a way that generates only a mild effect on the light curve (and therefore might not be easily detected). We find that in such a case the errors of our model, which assumes constant microphysics, are  comparable to the errors  induced by different jet structures. We conclude that as long as the  microphysical parameters evolve in a way that is not easily seen in the light curve, the errors they introduce will most likely do not have a major effect on the uncertainty in the measurement of the jet geometry. 

{\bf GW170817}: Applying our model to the observations of GW170817 we find $\theta_{obs}=19.4 \pm 2.1 ~\deg$ ($1\sigma$ uncertainty), where the error includes the uncertainties in the image displacement, host galaxy distance, values of $T_p$ and $T_{end}$ and the uncertain jet structure, where the first two sources in this list dominates the uncertainty. We find that the jet opening angle at the time of the peak is $\theta_c=1.5-4~\deg$, where the main source of uncertainty is the identification of $T_p$ and $T_{end}$. Our measurements agree with those of \cite{Mooley2022} for the value of $\theta_{obs}-\theta_c$ but our estimate of $\theta_{obs}/\theta_c$ is larger. As a result we obtain a smaller value for $\theta_c$  and a slightly lower value for $\theta_{obs}$. We discuss possible sources for this mild disagreement in the main text. 

Finally, our main conclusion is that measurements of the image displacement and/or the image shape enable measurement of $\theta_{obs}$ and $\theta_c$ rather accurately. However, only a small subset of entire sample of future GW detected BNS and BH-NS mergers are expected to have such observations. The reason is that the afterglow is expected to be bright enough for such measurements only when the jet axis is not too far away from the line-of-sight and the merger distance is not too high (the exact distance depends on the jet energy, circum-merger density and $\theta_{obs}$). Yet, this sub-sample can teach us a great deal about the physics of the merger, and possibly also improve the estimate of $H_0$. An interesting possibility is that scintillation will be observed in some of the mergers where the radio afterglow is detected but the image displacement cannot be resolved. This will provide constraints image size, which we plan to study in a future work. 

\section*{Acknowledgements}
We thank E. H. Ayache, B. D. Metzger, D. Kushnir and L. Sironi for the useful discussions and comments. This research was partially supported by a consolidator ERC grant 818899 (JetNS) and by an ISF grant (1995/21). TGS thanks the Buchman foundation for their support.  

\section*{Data Availability}
The data underlying this article will be shared on reasonable request to the corresponding author.

\bibliographystyle{mnras}
\bibliography{citations}

\appendix

\section{Derivation of $y_{max}$ For Power-Law Jets}\label{Appendix:power-law}
Consider a jet as described in \S\ref{sec:model}, with a structure:
\begin{equation}
    \frac{dE}{d\Omega} \propto \begin{cases}
1 & \theta\le\theta_{c}\\
\left(\frac{\theta}{\theta_{c}}\right)^{-b} & \theta>\theta_{c}.
\end{cases}
\end{equation}
To find the Lorentz factor dominating $y_{max}$, we must solve $\frac{d y(\xi,\psi=0)}{d\xi}=0$ at a constant observer time. First, let us find $y(\xi,\psi=0,T)$. By plugging Eq. \eqref{eq:R} into Eq. \eqref{eq:y_obs} we obtain (assuming $\xi\ll1$, $\Gamma\gg1$):
\begin{equation}\label{eq:y(xi)}
    y(\xi,\psi=0,T,\Gamma)\simeq\frac{c T \xi}{\frac{\xi^2}{2}+\frac{1}{2(2\omega+1)\Gamma^2}}~.
\end{equation}
This relation depends on $\Gamma$ and must be solved alongside an equation which relates $\Gamma$ to $T,\xi$ (along $\psi=0$). By assuming that every point on the shock evolves as $\Gamma\propto r^{-\omega}$, and using Eq. \ref{eq:R} to move to the observer frame we find:
\begin{equation}\label{eq:T(Gamma,PL)}
T=T_{0}\left(\frac{\Gamma}{\Gamma_{c}}\left(\frac{\theta_{obs}-\xi}{\theta_{c}}\right)^{b}\right)^{-\frac{1}{\omega}}\left(\frac{\frac{1}{\Gamma^{2}\left(2\omega+1\right)}+\xi^{2}}{\frac{1}{\Gamma_{c}^{2}\left(2\omega+1\right)}+\left(\text{\ensuremath{\theta_{obs}}}-\theta_{c}\right)^{2}}\right)~,
\end{equation}
 where $\Gamma_c$  and $T_0$ are constants set such that at time $T_0$, the closest part of the core (i.e, $\xi=\theta_{obs}-\theta_c$) the Lorentz factor will be $\Gamma_c$. 
 Next, we derive $\frac{d y(\xi,\psi=0)}{d\xi}$ by the chain rule, holding $T$ constant:
\begin{equation}\label{eq:dydxi}
    \frac{d y}{d\xi}=\frac{\partial y}{\partial \xi}+\frac{\partial y}{\partial \Gamma}\frac{d \Gamma}{d \xi}
    = -2\left(2\omega+1\right) c T \Gamma\frac{\Gamma^{3}\xi^{2}\left(2\omega+1\right)-\Gamma-2\frac{d\Gamma}{d\xi}\xi}{\left(\Gamma^{2}\xi^{2}\left(2\omega+1\right)+1\right)^{2}}~.
\end{equation}
We find $\frac{d \Gamma}{d \xi}$, by implicitly deriving Eq. \eqref{eq:T(Gamma,PL)} for constant $T$ and solving for $\frac{d\Gamma}{d\xi}$,
\begin{equation}\label{eq:dGammadxi}
\frac{d\Gamma}{d\xi}=\Gamma\frac{\left(2\Gamma^{2}\left(2\omega+1\right)-1\right)\left(b\cos\xi-\omega\left(\theta_{obs}-\xi\right)\sin\xi\right)-2b\Gamma^{2}\left(2\omega+1\right)}{\left(2\omega+1\right)\left(\theta_{obs}-\xi\right)\left(\left(2\Gamma^{2}-1\right)\cos\xi-2\Gamma^{2}\right)}~.
\end{equation}
plugging it into \eqref{eq:dydxi}, we obtain
\begin{equation}
        \frac{d y}{d\xi}=2c\Gamma^{2}T\frac{2b\xi-\theta_{obs}\left(2\omega+1\right)\left(\Gamma^{2}\xi^{2}-1\right)+\xi\left(2\omega+1\right)\left(\Gamma^{2}\xi^{2}-1\right)}{\left(\Gamma^{2}\xi^{2}+1\right)\left(\theta_{obs}-\xi\right)\left(\Gamma^{2}\xi^{2}\left(2\omega+1\right)+1\right)}.
\end{equation}
Setting the nominator equal to zero, we find: 
\begin{equation}\label{eq:Gamma_edge}
\Gamma^{max} = \frac{1}{\xi^{max}}\sqrt{\frac{\frac{b\xi^{max}}{\theta_{obs}-\xi^{max}}+2\omega+1}{2\omega+1} }~,
\end{equation}
Where $\Gamma^{max}$ is the shock Lorentz factor at the point  $(\xi,\psi)=(\xi^{max},0)$. To find $y_{max}$ explicitly, \eqref{eq:Gamma_edge} can be plugged into \eqref{eq:y(xi)} and \eqref{eq:T(Gamma,PL)} to give a parametric solution $(y_{max}(\xi^{max}),T(\xi^{max}))$. Note that equation \eqref{eq:Gamma_edge}, $\Gamma^{max}\xi^{max}>1$ for all $\xi^{max}>0$ and $b>0$, while as expected, $\Gamma\xi=1$ for a sphere $(b=0)$. This means that once $\xi^{max}=\theta_{obs}-\theta_c$, and we would expect to start seeing into the core, the observer does not immediately start seeing deeper into the core, but rather $y_{max}$ joins the evolution of a top-hat jet before the time of the peak. The larger $\Gamma^{max}(\theta_{obs}-\theta_c)$ is, the earlier $y_{max}$ joins the evolution of a top-hat jet. This means that for steeper jet structures (larger $b$) and for larger values of $\frac{\theta_{obs}}{\theta_c}$ the evolution joins that of a top-hat jet at an earlier observer time. A few examples of these solutions are plotted in Fig. \ref{fig:power-law}, and this behaviour can be seen there.

\section{Calibrated relations for the depth and width of the image }\label{Appendix:depth} 
\subsection{The depth of the image}
To derive the depth of the image, it is useful to recall the description of the image as an intersection of a circle and a ring, as seen in Fig. \ref{fig:ring_jet_intersection}. 
Consider a ring with radius $R_{arc}=r\xi$ and thickness $\sim 0.18 R_{arc}$  (the thickness of the ring is discussed briefly in \S\ref{sec:sphere}) intersecting a circle with radius $r \theta_c$. In most cases the depth of the image emerging from the intersection is a combination of the curvature of the arc, $\Delta y_{curve}=(1-\cos\frac{\theta_c}{\theta_{obs}})R_{arc}$, and the thickness $\Delta y_{thickness}\sim 0.18 R_{arc}$. However, if $\theta_{obs}\gg \theta_{c}$, we will not see the full thickness of the ring, and the depth will be determined by the size of the core, and be comparable to the width: $\Delta y\simeq \Delta x \simeq 2\cdot y_{cen}\frac{\theta_c}{\theta_{obs}}$. Thus, we expect that always $\frac{\Delta y}{\Delta x}\lesssim1$, where this limit is approached at large $\frac{\theta_{obs}}{\theta_c}$. To conclude, the depth is a combination of three effects (image curvature and thickness and the core size), where the importance of each effect varies between different cases.
In the range of $\frac{\theta_{obs}}{\theta_c}$ used in this work, we probe mainly the transitional phase between the different regimes, and none of these effects can be neglected. We find that in the range of $\theta_{obs}/\theta_c$ that we explore numerically, $\frac{\Delta y}{\Delta x} $ can be reasonably approximated as linear in $\frac{\theta_{obs}}{\theta_c}$ and $\frac{\Delta y}{y_{cen}} $ can be reasonably approximated as linear in $\frac{\theta_c}{\theta_{obs}}$ . 
\subsection{Calibration of relations for the depth and width}
As discussed in \S\ref{sec:Width}, the width can be related to the centroid location by:
\begin{equation}\label{eq:width_calib}
    \Delta x= C_{width}\frac{\theta_{c}}{\theta_{obs}}y_{cen}~,
\end{equation}
where $C_{width}$ is a calibration constant and we expect $C_{width}\simeq 2$. 

As seen in Fig. \ref{fig:depth_over_width}, the depth of the image can be written as:
\begin{equation}\label{eq:depth-width}
    \Delta y = (C_{dw1}+C_{dw2} \frac{\theta_{obs}}{\theta_{c}})\Delta x~.
\end{equation}
From these two relations, we can write an expression for the relation between the depth and the centroid:
\begin{equation}\label{eq:depth-cen}
        \Delta y = (C_{dc1}\frac{\theta_c}{\theta_{obs}}+C_{dc2}) y_{cen}~.
\end{equation}
The calibration constants for all these expressions are listed in table \ref{tab:depth_width_calib}.
Note that since the expressions for $\Delta y$ do not capture the asymptotic behaviour, they should not be extrapolated to larger values of $\frac{\theta_{obs}}{\theta_c}$ than those that we explore here numerically.  

\begin{table*}
    \centering
\begin{tabular}{|c|c|c|c|c|c|c|}
\hline 
\multirow{2}{*}{} & \multirow{2}{*}{Calibrtion Constant} & \multicolumn{5}{c|}{$p$}\tabularnewline
\cline{3-7} \cline{4-7} \cline{5-7} \cline{6-7} \cline{7-7} 
 &  & 2.05 & 2.2 & 2.5 & 2.8 & 3\tabularnewline
\hline 
Depth-Width, eq. \ref{eq:depth-width} & $C_{dw1}$ & 0.26 & 0.29 & 0.35 & 0.41 & 0.46\tabularnewline
 & $C_{dw2}$ & 0.06 & 0.05 & 0.04 & 0.03 & 0.03\tabularnewline
Depth-Centroid, eq. \ref{eq:depth-cen} & $C_{dc1}$ & 0.55 & 0.51 & 0.53 & 0.57 & 0.59\tabularnewline
 & $C_{dc2}$ & 0.14 & 0.14 & 0.13 & 0.12 & 0.12\tabularnewline
\hline 
\multirow{1}{*}{Width-Centroid, eq. \ref{eq:width_calib}} & $C_{width}$ & 2.2 & 2.17 & 2.12 & 2.08 & 2.06\tabularnewline
\hline 
\end{tabular}
    \caption{Calibration constants for the depth-width (Eq. \ref{eq:depth-width}), depth-centroid (Eq. \ref{eq:depth-cen}) and width-centroid (Eq. \ref{eq:width_calib}) relations.}
    \label{tab:depth_width_calib}
\end{table*}

\section{Numerical Setup and Convergence Tests}\label{Appendix:Numerical_Setup}

In all our simulations with GAMMA we use piece-wise linear spacial reconstruction, hllc solver and third order Runge-Kutta time stepping and a CFL of 0.4. 
The grid  setup is similar to the one described in section 5 of \cite{Ayache2022}. It has 2D spherical coordinates $(r,\theta)$, with axial symmetry along the jet axis, where in the $\theta$ direction the grid boundaries are $[0,\pi/2]$. We chose the grid angular spacing such that there is high angular resolution in the region in which there is a relativistic shock, and lower resolution at larger angles. The track boundaries in the $\theta$ direction are given by
\begin{equation}\label{eq:grid}
    \theta_{j-\frac{1}{2}}=\frac{\pi}{2}\left(f_R \frac{j}{N_\theta}+\left(1-f_R\right)\left(\frac{j}{N_\theta}\right)^\alpha \right)~,
\end{equation}
where in all our simulations $\alpha=5$, $N_\theta$ is the number of tracks, set so the width of the track along the axis is $\Delta\theta\simeq\frac{1}{5\gamma_0}$, and $f_R$ is the fraction of the grid in which we expect to see an ultra-relativistic shock. The parameters used in every simulation are summarized in table \ref{tab:grid_setup}.
In the $r$ direction, the AMR sets the resolution. The re-gridding scheme is set to runaway in order to fully resolve the shock. We set the re-gridding score as $S_{regrid}=\frac{\Delta r}{r_{max} \Delta \theta}\gamma^{3/2}$, and allow $S_{regrid}$ to vary in $[0.1,3]$ while $\gamma_{shock}(\theta=0)\ge 30$, and $[0.2,6]$ after that. In addition, the resolution in the 10 cells ahead of the shock is increased by a factor of 10. 

We initialize the simulation by setting the initial conditions in each track as part of a Blandford-Mckee solution with energy $E_{iso}(\theta)$ and uniform cold external density; $P=10^{-5}\rho c^2$. The grid boundaries in the radial direction are initially $[R_{shock,0}(1-\frac{50}{\Gamma_0^2}),R_{shock,0}(1+\frac{50}{\Gamma_0^2})]$, where $R_{shock,0}$ is the radius of the part of the shock with the highest Lorentz-factor, and $\Gamma_0$ is the highest shock Lorentz factor. The outer radial boundary moves so that it is always a bit ahead of the shock - $r_{max}=1.1 c t$ and the inner boundary does not move. 

We start with 5000-7000 cells in each track in the radial direction in order to resolve the shock, and let the re-griding scheme reduce the resolution to $\sim 1000$ within a few time steps.
We use reflective boundary conditions at the inner radial boundary, for the sake of energy conservation. It turns out that simulations with a reflective inner radial boundary also run faster than an outflow boundary, since an outflow boundary causes many new cells to form in regions with negligible energy. We use outflow conditions at the outer radial boundary so that new external medium will enter the computation domain as the shock advances, and outflow at the outer angular boundary, since the shock doesn't reach the outer boundary in the course of the simulation, so the boundary condition doesn't matter. 

\begin{table}
    \centering
\begin{tabular}{|c|c|c|c|c|}
\hline 
\multicolumn{2}{|c|}{Simulation} & $\theta_{c,0}$ & $N_{\theta}$ & $f_{R}$\tabularnewline
\hline 
\hline 
\multicolumn{1}{|c}{\multirow{2}{*}{Top-Hat}} & \multirow{2}{*}{} & $0.05$ & 200 & 0.15\tabularnewline
\cline{3-5} \cline{4-5} \cline{5-5} 
 &  & $0.15$ & 300 & 0.225\tabularnewline
\hline 
\multirow{5}{*}{Power-Law} & \multirow{1}{*}{$b=3$} & $0.05$ & 500 & 0.4\tabularnewline
\cline{2-5} \cline{3-5} \cline{4-5} \cline{5-5} 
 & \multirow{2}{*}{$b=6$} & $0.05$ & 200 & 0.15\tabularnewline
\cline{3-5} \cline{4-5} \cline{5-5} 
 &  & $0.15$ & 500 & 0.5\tabularnewline
\cline{2-5} \cline{3-5} \cline{4-5} \cline{5-5} 
 & \multirow{2}{*}{$b=12$} & $0.05$ & 200 & 0.15\tabularnewline
\cline{3-5} \cline{4-5} \cline{5-5} 
 &  & $0.15$ & 300 & 0.225\tabularnewline
\hline 
\multicolumn{1}{|c}{Gaussian} & \multicolumn{1}{c|}{} & $0.065$ & 200 & 0.15\tabularnewline
\hline 
\multirow{2}{*}{Hollow} & \multicolumn{1}{c|}{Top-Hat} & $0.05$ & 200 & 0.15\tabularnewline
\cline{2-5} \cline{3-5} \cline{4-5} \cline{5-5} 
 & $b=6$ & $0.15$ & 400 & 0.4\tabularnewline
\hline 
\end{tabular}
    \caption{The grid parameters defined in Eq. \eqref{eq:grid} are listed for each of the simulations}
    \label{tab:grid_setup}
\end{table}

\subsection{Convergence Tests}\label{Appendix:Convergance}
\begin{figure}
    \centering
    \includegraphics[width=0.5\textwidth]{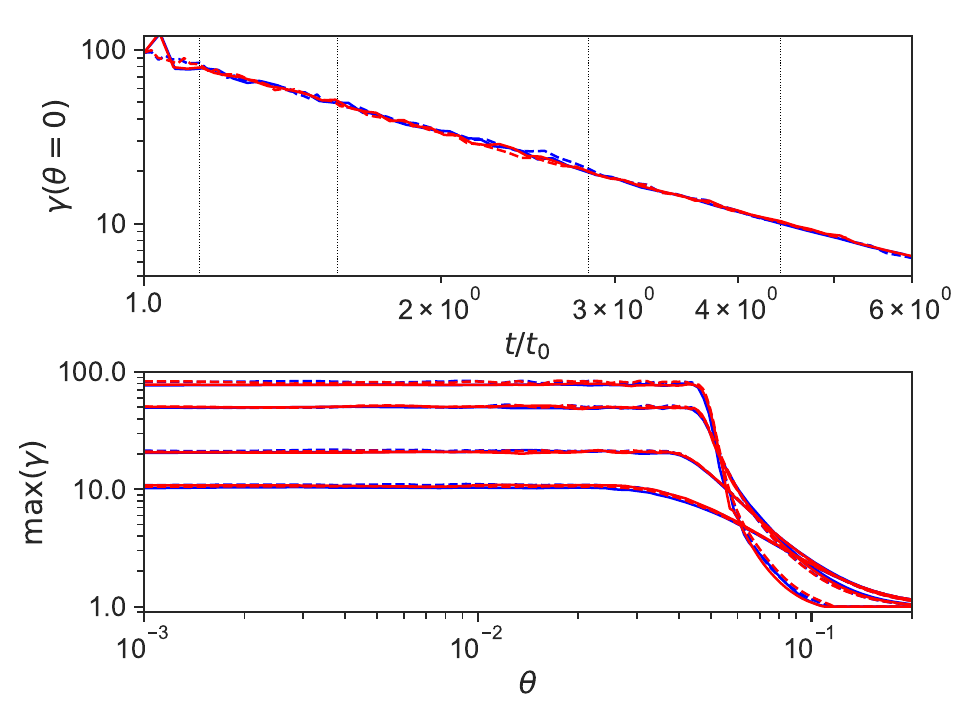}
    \caption{The top panel shows the highest matter Lorentz factor along the axis, as a function of the lab time. The bottom panel shows the angular matter Lorentz factor structure along the shock front as a function of angle for several lab times, these times are marked in black dotted lines in the top panel. In both panels, the dashed lines denote simulations with $N_\theta=400$ and the solid line denotes simulations with $N_\theta=200$. The blue lines are for $\alpha =3$ and the red ones are for $\alpha =5$.}
    \label{fig:conv_test}
\end{figure}

We ran convergence tests to check the grid resolution, the grid setup and the re-griding scheme. For the convergence tests we ran simulations of a top-hat jet, since the sharp cutoff makes it the most difficult structure to resolve correctly. We ran 4 simulations with an opening angle of 0.05 and initial Lorentz factor of 100, and a grid in the $\theta$ direction given by Eq. \ref{eq:grid} with $f_R=0.15$. In two of these simulations $\alpha=3$ (like in \citealt{Ayache2022}) and $N_{\theta}=200,400$, and in the other two $\alpha=5$ (as in our simulations) and  $N_{\theta}=200,400$ (note that thanks to the regridding scheme, an increase in the resolution in the $\theta$ direction contributes to a comparable increase in the resolution in the $r$ direction). These 4 simulations produced indistinguishable jet structures ($E_{iso}(\theta)$).
For the simulation with a $N_{\theta}=200$, $\alpha=5$, we compared changing the range of  $S_{regrid}$ from $[0.1,3]$ to $[0.2,6]$ when the shock is at a Lorentz factor of 30, and leaving it $[0.1,0.3]$ also once the shock decelerates. We compared the outcome of these simulations by looking at the shock Lorentz factor evolution at different angles, and found all these setups in good agreement. 
Fig. \ref{fig:conv_test} shows a comparison of the different grid setups. The simulations seem indistinguishable up to random numerical noise. The comparison of simulations with different treatment of $S_{regrid}$ are identical in the same manner. 

\subsection{Comparison to PLUTO}\label{Appendix:PLUTO}
Using PLUTO without AMR, or any simulation with a static grid without AMR, to simulate ultra-relativistic jets is challenging, since extremely high resolution is required in order to resolve the shock. In the simulations with PLUTO we were not able to resolve shocks with $\Gamma\ge 5-6$. However, we were able to compare the energy profile evolution to the one in GAMMA.
\subsubsection{PLUTO Simulation Setup}
We used the 2D RHD module on PLUTO 4.4 with spherical geometry, a static grid, hll solver, CFL of 0.2, parabolic reconstruction and 3rd order Runge-Kutte time stepping. Like in the GAMMA simulations, we used an ideal equation  of state, with an adiabatic index of $4/3$. 
The grid consisted of 8685 uniformly distributed cells in the $\theta$ direction, $0\le\theta\le 1$, and the main grid in the $r$ direction spans $1\le r\le 15$ with 23,800 cells logarithmically spaced. The grid in the $r$ direction is extended by 700 cells logarithmically decreasing in space between $0.9$ and 1, used to set the initial conditions. The boundary conditions are reflective at the inner $r$ boundary, axisymmetric at $\theta=0$, and outflow at the other two boundaries.

\subsubsection{Comparing GAMMA and PLUTO Simulations}
\begin{figure}
    \centering
    \includegraphics[width = \columnwidth]{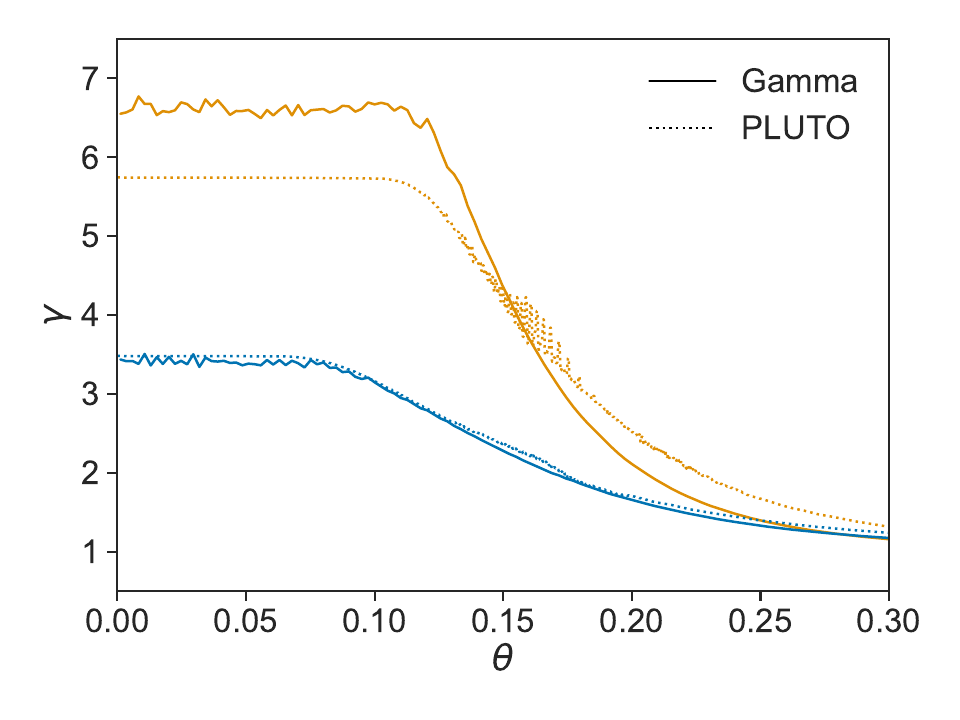}
    \caption{The profile of $\gamma(\theta)$ for GAMMA and PLUTO top-hat simulations is compared at two lab times. In the first, the PLUTO simulation is not yet converged while in the second the two simulations are very similar.}
    \label{fig:GAMMA_PLUTO_comp}
\end{figure}

In the PLUTO simulations, the matter Lorentz factor is not converged until the Lorentz factor is $\gamma\simeq 4$. For this reason, we could not use PLUTO simulations to study the afterglow properties. However, we can compare the eventual angular shock structure. 
We compared simulations of a top-hat jet with an initial matter Lorentz factor of 20, and $\theta_{c,0}=0.15$.
In Fig. \ref{fig:GAMMA_PLUTO_comp} we compare the Lorentz factor of the matter along the shock as a function of $\theta$ of the two simulations at two times. In the first, the PLUTO simulation is not yet converged while in the latter both simulations follow an identical angular structure. 

\section{Post-Processing Code}\label{Appendix:post_processing}
In the post-processing code, we calculate for each RHD simulation the images and light curve in the power-law segment $\nu_a,\nu_m<\nu_{obs}<\nu_c$ as seen by observers at different observing angles.
We construct 3D grids with all the physical fluid parameters from the 2D simulations by rotating along the jet axis. For every lab time, we scan through the mass elements in every lab frame,  and derive their contribution to observer times and locations on the sky plane. The flux contributed by each cell is calculated by following the standard afterglow model \citep{Sari_piran_Narayan1998}.

By the definition of $\epsilon_B$, the magnetic field in the shocked fluid is given by:
\begin{equation}
    B=\sqrt{24 \pi \epsilon_B P}~,
\end{equation}
where $P$ is the pressure. 
The minimal Lorentz-factor of the power-law electron distribution is:
\begin{equation}
    \gamma_m=\epsilon_e\frac{p-2}{p-1}\frac{3 P m_p}{m_e c^2 \rho}~,
\end{equation}
and the typical synchrotron frequency is:
\begin{equation}
    \nu_m = \frac{q_e}{2\pi m_e c}\gamma_m^2 B~.
\end{equation}
The cell specific luminosity at the peak of $L_\nu$ is given by
\begin{equation}
    \Delta L_{\nu,max} = \frac{m_e c^2 \sigma_T B}{3 q_e m_p}\cdot \gamma \rho r^2\Delta r \sin \theta \Delta \theta \Delta \phi~,
\end{equation}
where $\gamma\rho$ is the rest frame density, multiplied by the cell volume to give the cell rest-mass. 
To transform to the observer frame we define the Doppler factor
    $\delta_D = \frac{1}{\gamma\left(1-\beta\cos\xi_v\right)}$ ,
where $\xi_v$ is the velocity angle to the observer, analogous to $\xi$.
The contribution of this cell to the observed frame flux is
\begin{equation}
    \Delta F_\nu = \Delta L_{\nu,\max}\left(\frac{\nu_{obs}\cdot (1+z)}{\nu_m\delta_D} \right)^\frac{1-p}{2}\cdot \frac{\delta_D^3}{4\pi D_L^2}
    \end{equation}
where $z$ is the red-shift and $D_L$ is the luminosity distance. 
This flux is added to the relevant cell in the image at the relevant observer times.

\section{Numerical Examination of Assumptions Used in the Analytical Model}\label{Appendix:Assumptions}
In the analytic model, we use various assumptions. The comparison of the model to numerical simulations shows that our analytic model provides a good description up to calibration constants of order unity. However, in order to better understand the system, we examine numerically two of the assumptions used in the model. 

\subsection{The Region Dominating the Observed Flux}\label{Appendix:Gamma_xi}
\begin{figure}
    \centering
    \includegraphics[width = \columnwidth]{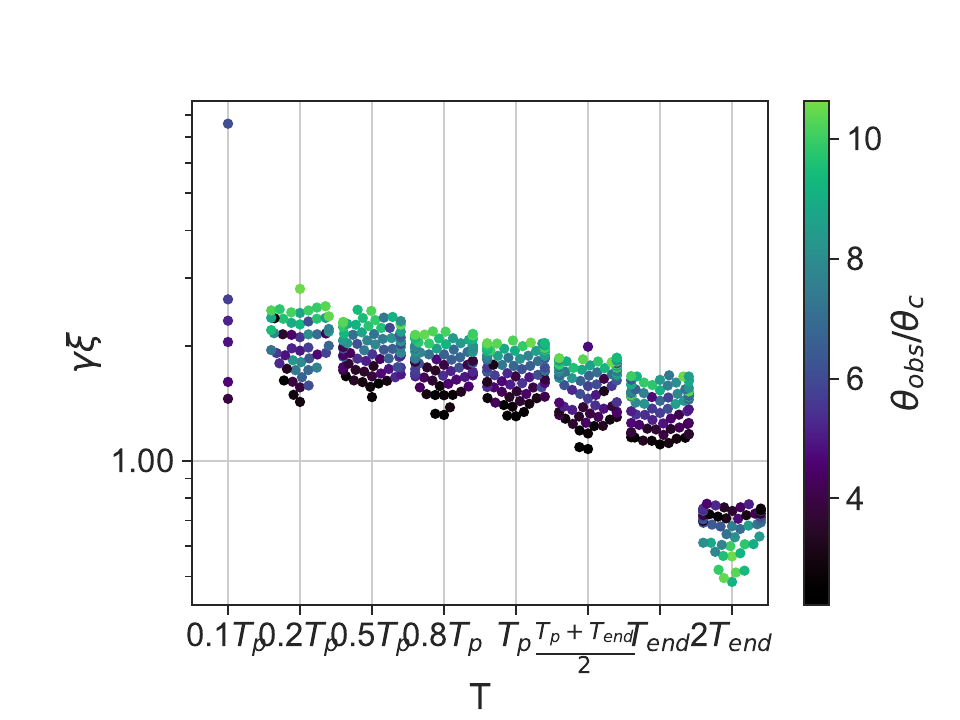}
    \caption{The flux-averaged value of $\gamma\xi$ at the brightest point in the image.
    }
    \label{fig:gamma_xi_swarm} 
\end{figure}

A key point in the analytic model is that for a significant part of the light curve, and especially around the peak, the region dominating the emission satisfies $\Gamma\xi \sim 1$.  At early times, during the rising phase when the structure surrounding the core dominates the emission, the emitting region is expected to satisfy $\Gamma\xi> 1$, where the exact value depends on the jet structure and on $\theta_{obs}/\theta_c$. In this phase, the emitting region of jets with sharper wings should have larger values of $\Gamma\xi$. As the light curve approaches the peak for all jet structure we expect $\Gamma\xi \simeq 1$. In Fig. \ref{fig:gamma_xi_swarm}, we see the value of $\gamma\xi$ (calculated as the flux-averaged $\gamma\xi$), where $\gamma$ is the matter Lorentz factor at the brightest point on the image for each of our simulations at different times. This figure shows that the analytic prediction roughly holds.

Before the peak, both the value of $\gamma\xi$ and the dispersion in this value  are larger, and both decrease as $T\to T_p$. In fact, at $0.1T_p$ one simulation (of a top-hat jet) even has $\gamma\xi\simeq8$. We find that the value of $\gamma\xi$ during the peak is slightly larger than 1 and more interestingly, that $\gamma\xi$ slightly decreases during the peak phase. The decrease in $\gamma\xi$ is mainly a geometrical effect - the energy distribution in the core in not completely flat, and the steeper it is, the larger the value of $\gamma\xi$ (as seen in appendix \ref{Appendix:power-law}). Both these effects are treated by the calibration coefficint(s) of the centroid motion. At $T>T_{end}$, $\gamma\xi<1$ as expected. 
One may notice that from $T\simeq 0.8T_p$ the value of $\gamma\xi$ is correlated with $\theta_{obs}/\theta_c$, this is probably a combination of the geometrical effect described earlier and of spreading. 

\subsection{Comparing Centroid and Edge Location}\label{Appendix:Centroid_Edge}
One of the approximations used in the analytical section was that the centroid location proportional and close to the location edge of the image. 
\begin{figure}
    \centering
    \includegraphics[width= \columnwidth]{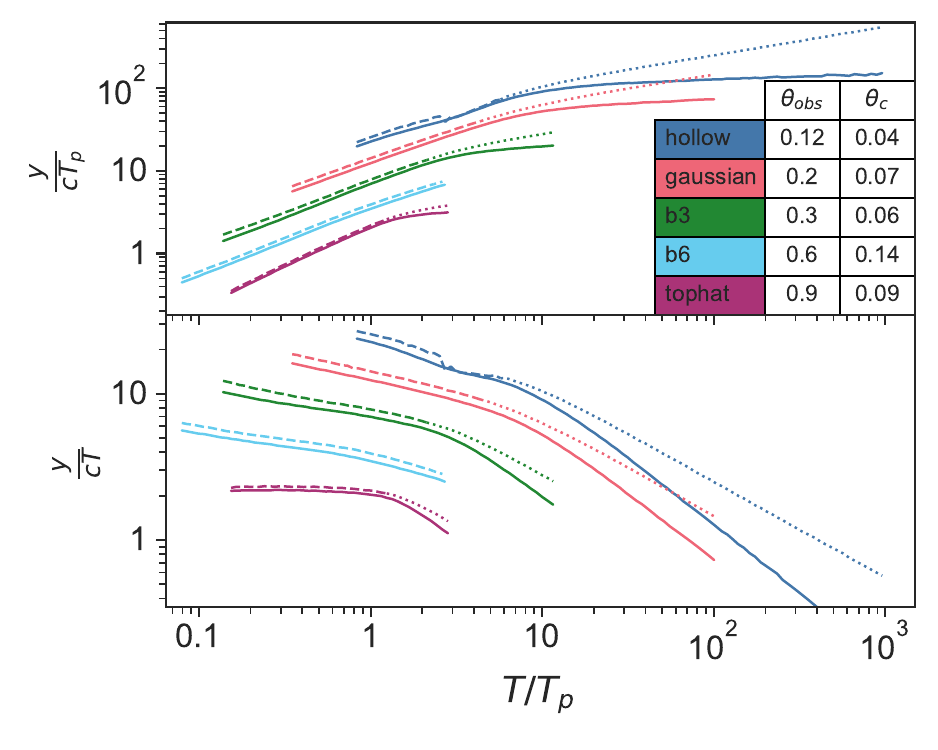}
    \caption{In the top panel, the centroid location $y_{cen}$ (solid line) and edge location $y_{max}$ (dashed line for $ T\le T_{end}$, dotted for later times) are compared for several sets of parameters. In the bottom panel, the average velocity of the centroid and edge is compared for the same simulations. In both panels it can be seen that until $T_{end}$, $y_{max}$ is larger than $y_{cen}$ by about $10\%$, and they differ significantly from each other only after $T_{end}$ when the emitting region is much more extended.}
    \label{fig:edge_vs_centroid}
\end{figure}
In Fig. \ref{fig:edge_vs_centroid}, the edge and centroid location are compared for several simulations. In all the simulations, between $0.2 T_p$ and $T_{end}$ the centroid is slightly behind the edge, as expected. Simulations of jets with a wide structure, like those with a power-law jet with $b=3$, lead to more extended emission and a larger deviation between the centroid and edge location. 

In all our simulations, we find that between $0.5T_p$ and $T=T_{end}$, the centroid location is very close to the edge, at $85\%-95\%$ of the edge location, with very little dispersion between simulations. At earlier times, the dispersion is slightly larger, the centroid location is $80\%-100\%$ of the edge location, and the relation depends on the jet structure and $\theta_{obs}/\theta_c$.

\end{document}